\RequirePackage[2020-02-02]{latexrelease}
\documentclass[11pt,nofootinbib,aps,letterpaper,preprintnumbers,superscriptaddress,showpacs]{revtex4}
\usepackage{amsmath}
\usepackage{eurosym}
\usepackage{amssymb}
\usepackage{graphicx}
\usepackage{color}
\usepackage{braket}
\usepackage{amsmath}
\usepackage{amssymb}
\usepackage{subfigure}
\usepackage{amsfonts}
\usepackage{natbib}
\usepackage[left=2cm,right=2cm,top=1.5cm,bottom=1.5cm]{geometry}
\usepackage[usenames,dvipsnames,svgnames]{xcolor}  %% colored text
\usepackage{hyperref}   %% Needs to make hyper reference of any references or citations
\definecolor{beamer@PRD}{RGB}{46,48,146}

\begin{document}
\newcommand\be{\begin{equation}}
\newcommand\ee{\end{equation}}
\newcommand\bea{\begin{eqnarray}}
\newcommand\eea{\end{eqnarray}}
\newcommand\bseq{\begin{subequations}} %solo con amsmath
\newcommand\eseq{\end{subequations}}
\newcommand\bcas{\begin{cases}}
\newcommand\ecas{\end{cases}}
\newcommand{\p}{\partial}
\newcommand{\f}{\frac}
%%%%%%%%%%%%%%%%%%%%%%%%%%%%%%%%%%%%%%%%%%%%%%%%%%%%%%%%%%%%%%%%%%%%%

\title{Emergent Universe from Energy-Momentum Squared Gravity}

\author {\textbf{Mohsen Khodadi}}
\email{m.khodadi@ipm.ir}
\affiliation{School of Astronomy, Institute for Research in Fundamental Sciences (IPM)\\
P. O. Box 19395-5531, Tehran, Iran}
\author {\textbf{Alireza Allahyari}}
\email{alireza.al@ipm.ir}
\affiliation{School of Astronomy, Institute for Research in Fundamental Sciences (IPM)\\
P. O. Box 19395-5531, Tehran, Iran}
\author{\textbf{Salvatore Capozziello}}
\email{capozziello@na.infn.it. Corresponding author.}
\affiliation{Department of Physics ``E. Pancini'', University of Naples ``Federico II'', Naples, Italy.}
\affiliation{INFN Sez. di Napoli, Compl. Univ. di Monte S. Angelo, Edificio G, Via	Cinthia, I-80126, Naples, Italy.}
\affiliation{Scuola Superiore Meridionale, Largo San Marcellino 10, I-80138, Naples, Italy.}
\affiliation{Tomsk State Pedagogical University, ul. Kievskaya, 60, 634061 Tomsk, Russia.}

\date{\today}
 %---------------------------------------

\begin{abstract}
In order to bypass the big bang  singularity, we develop an emergent universe  scenario
within a covariant  extension of General Relativity known as \emph{``Energy-Momentum Squared Gravity''}. The extra terms  of  the model emerge in the high energy regime.  Considering dynamics  in a Friedmann-Lema\^itre-Robertson-Walker background, critical points, representing stable Einstein static states of the phase space, result as solutions.
It then turns out that as the equation of state parameter $\omega$ gradually declines from a constant value
as $t\rightarrow-\infty$, eventually some of the static past eternal solutions find the chance to naturally enter into thermal history through a graceful exit mechanism. In this  way, the successful
realization of the emergent universe allows an  expanding thermal history without the big bang singularity
for the spatially flat universe free of cosmological constant.
%In principle, this situation can be compared with the observations.
% contrary to the standard EU scenario.

\end{abstract}
\maketitle

\section{Introduction} \label{Int}
Although  inflationary paradigm seems to be sufficiently supported by  data so that one can take it as the  default approach to describe the early universe,  the initial singularity issue (the so-called Friedmann-Lema\^itre-Robertson-Walker (FLRW) big bang) is far from being clear and definitely solved in the framework of this paradigm. On the basis of some theorems \cite{Book}, such singularities
are generic and unavoidable, meaning that the classical spacetime has a threshold point  beyond which the standard
general relativity (GR) is not applicable \cite{Borde:1996pt, Borde:2001nh}. The singularities are commonly diagnosed
by divergences of the scalar invariants of  curvature or torsion  tensors or the collapse of geodesics at some given points. Therefore, it seems that the initial
singularity issue will potentially shed light on the answer to the  question of whether our universe has a beginning
or  eternally existed. The lack of a solution for this fundamental issue in the context of inflationary cosmology\footnote{It is worth  recalling two points. First of all, the inflationary scenario is in conflict with singularity theorems by Penrose and Hawking, due to explicit violation of strong energy condition. Secondly, inflation cannot be  eternal in the past  since it suffers from geodesic incompleteness issue. For a detailed discussion about the Pros and Cons of inflationary cosmology, see \cite{Brandenberger:2010dk, Brandenberger:2012um} and also \cite{Penrose:1988mg}.}  motivated many authors to construct
some pre-inflationary scenarios such as: emergent universe (EU) \cite{Ellis:2002we, Ellis:2003qz}, cyclic/ekpyriotic scenarios
\cite{Khoury:2001wf, Steinhardt:2001st, Khoury:2003rt, Barrow:2004ad} which  are commonly non-singular or past eternal.
Motivations by the widespread belief that including a quantum  gravity (QG) effects at very short scales leads to the natural disappearance of
singularities has led to some other singularity free cosmological models \cite{Bojowald:2001xe, Hossain:2009ru, Battisti:2009zzb, Alesci:2016xqa}.
Those models  have been merely derived from semi-classical  corrections to QG, see \cite{Ali:2014hma, Vakili:2015wbw, Khodadi:2016bcx},  or considering non-local corrections \cite{Bahamonde:2017sdo,Capozziello:2020xem} as well.

The so-called EU scenario is one of the popular candidates that has been highly considered by several authors.
This scenario, which is equivalent to \emph{``a creation in  absence of beginning of time''},
includes striking traits: no initial singularity (no beginning of time i.e infinite past), no horizon problem and no QG era. More exactly,
Ellis et al \cite{Ellis:2002we, Ellis:2003qz}  proposed a scenario to overcome the initial singularity within the framework
of GR. In this framework a closed\footnote{Despite that most data suggest the universe is effectively spatially flat, meaning it has no curvature similar to a sheet of paper,  recent Planck measurements do not exclude  that it could be closed \cite{DiValentino:2019qzk,Vagnozzi:2020rcz}.}
FLRW cosmology with a positive spatial curvature is preceded by
an initially static state known as the Einstein Static Universe (ESU)\footnote{It may be interesting to know that the idea of ESU as a vital component in EU scenario actually was originated in the seminal papers by Eddington and Lema\^itre, respectively \cite{Eddington:1930zz, Lemaitre:1931zza} .} in the eternal past (instead of a big bang singularity). Finally this  closed FLRW  is superseded by  the inflationary era. According to this scenario, the closed universe has existed eternally, but eventually, at
some point, it begins inflating \cite{Mukherjee:2006ds}. As a result, there are two required conditions for the scenario to provide a successful
description for fixing  the initial singularity:  a self-consistent exit from the ESU  i.e. stable as well as a graceful
exit from the ESU to the inflation. Note that the former and the latter  are respectively necessary and sufficient
conditions for the singularity circumvention so that  the failure in satisfying any of the two conditions will  cease the EU scenario.
Failure to meet the former condition caused this scenario to  face  a challenge in the first place.
The original EU scenario failed to solve the big bang singularity issue successfully since Barrow et al \cite{Barrow:2003ni}
discovered that the ESU in  GR is not stable, meaning that the universe in such an initial static state is not
able to survive for a long time against the existence of dominating perturbations. However,  the extreme physical
conditions expected in the early universe (such as those arising from quantization of gravity, or corrections to GR)
may turn the stability situation of initial state in favor of EU scenario. In other words, due to the failure of the EU scenario in the context of GR,  modified theories of gravity may have the potential to improve the situation.
This idea has led to several  studies on the natural
extensions of the original EU setup into modified gravitational theories with the aim to derive some promising outputs in comparison with GR, \cite{Boehmer:2003iv}-\cite{Sharif:2021}.

Recently, a modification of the matter Lagrangian (instead of gravitational Lagrangian) in a nonlinear way using a term proportional to $T_{\mu \nu }T^{\mu \nu }$ has been proposed as a new covariant generalization of GR \cite{Roshan:2016mbt, Board:2017ign}.
This theory is known as \emph{``energy-momentum squared gravity''} (EMSG) and induces quadratic contributions to gravity from matter side
(without the appearance of novel forms of fluid stresses such as the scalar field and so on \cite{Akarsu:2018zxl})
so that it affects the cosmological dynamics considerably at high energy phases. In other words, the self-coupling of matter, instead of geometry, is assumed having interesting cosmological consequences, in particular, at early epochs.
It is worth noticing  that the so-called EMSG actually is a special case of theories with the general form of the Lagrangian as $f(R, T_{\mu\nu}T^{\mu\nu})$ which were first investigated
in \cite{Arik:2013sti}.
Naturally, the expected deviations from GR in early universe
can lead to non-trivial consequences for some key issues in modern cosmology such as: initial singularity, inflation and big bang nucleosynthesis. Actually due to the lack of a final theory of QG, one of the motivations behind considering   modified gravities, such as EMSG, is trying to remove  the big bang singularity.
A  previous work \cite{Roshan:2016mbt} has shown  that this theory is singularity free due to the  bounce at early epochs. In other words, EMSG, by predicting a minimum length and a finite maximum energy density as common features of most QG approaches, cancels out the singularity issue of  standard GR scenario of early universe. It is remarkable that it is the incentive to work with the EMSG model and also its generalized versions as it is not restricted to discarding the initial singularity, but, depending on the underlying EMSG model, it may result in some interesting modifications to the whole cosmic history, too. Other motivation for using EMSG  come from trying to improve the usual paradigm in $\Lambda$CDM-based cosmology. Indeed, despite the $\Lambda$CDM is successful in fitting a wide range of the observational data, it not able to give a self-consistent  description of the cosmic acceleration  \cite{Capozziello:2011et}. Tensions in measurements of the universe acceleration between  early time and late time,  as the so-called  ``Hubble tension'', may be a robust  confirmation in the direction of modifying the $\Lambda$CDM picture as recently reported in  \cite{DiValentino:2021izs}.  In other words, one may imagine the EMSG and its generalized models as phenomenological extensions of the $\Lambda$CDM, such that, despite the existence of  a cosmological constant, the nonlinear matter Lagrangian, embedded in these models, leads to additional terms in Einstein's equations, which finally can be constrained via cosmological observations.
The cosmological applications of this novel modified theory of gravity along with its generalized types has attracted much attention in recent years (see some recent studies as \cite{Liu:2016qfx}-\cite{Shahidi:2021lqf}). In  \cite{Akarsu:2018zxl,Ranjit:2020syg},   the viability of EMSG has been studied via contrasting the relevant free parameter in light of  observations. In \cite{Akarsu:2018zxl},  authors  used the existing observational constraints on  the masses and radii of neutron stars and  derived some  constraints on the EMSG free parameter. In \cite{Ranjit:2020syg}, by adopting   recent observational data such  as  $i)$ cosmic chronometer and SNe Type-Ia Riess (292) $H(z)-z$ data-sets, $ii)$ baryon acoustic oscillation (BAO) peak parameter and $iii)$ cosmic microwave background (CMB) peak parameter,  bounds on the model parameter $\alpha$ have been derived. It is worth mentioning that, for  low redshift data, some constraints on the relevant model parameter of energy-momentum-powered gravity models have been achieved \cite{Faria:2019ejh}. Although that constraints obtained on EMSG free parameter are tight, still this theory is at the play.

The above descriptions motivated us to look for another cosmological implication of EMSG model i.e. the full realization of the EU scenario.
More exactly, we seek for the possibility whether EMSG model allows the EU scenario to be a viable solution of the early universe singularity. It is worth noticing that the study of the EU within the context of EMSG is well-motivated, because it changes the GR-based picture at early times, exactly when  the EU scenario comes into play.

This manuscript is structured as follows. In Sec.~\ref{sec:background},
we give a general review of the action as well as dynamical field equations
in the framework of EMSG model. In Sec.~\ref{ESU} ,we extract ESU solutions
related to the underlying gravity model and then investigate their stability
using first order dynamical system analysis. In Sec.~\ref{EC}, we look for a realistic
EU scenario via providing a graceful exit mechanism for the stable ESU solutions derived in
Sec.~\ref{ESU}.
Finally, we end the paper with a discussion summarized in Sec.~\ref{CD}.
Across this manuscript, for the signature of the spacetime metric, we will set
the $(-, +, +, +)$ convention.

%%%%%%%%%%%%%%%%%%%%%%%%%%%%%%%%%%%%%%%%%%%%%%%%%%%%%%%%%%%%%
\section{Energy-Momentum Squared Gravity and Background Field equations}\label{sec:background}

In EMSG model, the action can be written as  \cite{Roshan:2016mbt,Board:2017ign}
\begin{equation}\label{action}
S_{EMSG}=\frac{1}{2\kappa }\int \sqrt{-g}(R-2\Lambda +\alpha (T_{\mu \nu }T^{\mu \nu }))\
d^{4}x+\int L_{m}\sqrt{-g}\ d^{4}x,~~~ \kappa\equiv 8\pi G
\end{equation}
in which the Einstein-Hilbert action with a cosmological constant $\Lambda$ is
extended via adding a self contracting term of energy-momentum tensor (EMT), $T_{\mu\nu}T^{\mu\nu}$.
Here, $R$, $L_{m}$ and $\alpha$, respectively refer to the Ricci scalar,
the matter Lagrangian density and a real constant which addresses the gravitational coupling
strength of the underlying modification. The mass dimension of the EMSG parameter $\alpha$
is $[M]^{-6}$ and potentially can be any non-zero real value. So it is expected that, in early universe with high energy density, the EMSG is different from the standard GR so that, by going to lower energies,
this deviation has to disappear. The action (\ref{action}) can be re-express as follows
\begin{equation}\label{actionR}
S_{EMSG}=\int \sqrt{-g}\big(\frac{R-2\Lambda}{2\kappa} +L_{m,eff}\big)\ d^{4}x,~~~
L_{m,eff}=L_{m}+\frac{\alpha}{2\kappa}T_{\mu \nu }T^{\mu \nu }
\end{equation}
Varying the above action with respect to the
inverse metric, we acquire the following modified Einstein's field equations
\begin{equation}\label{Ef}
R_{\mu\nu}-\frac{1}{2}Rg_{\mu\nu}+\Lambda g_{\mu\nu}=\kappa T^{\text{eff}}_{\mu\nu}~,
\end{equation} where we have defined
\begin{eqnarray}\label{tef}
T^{\text{eff}}_{\mu\nu}&&=-\frac{2}{\sqrt{-g}}\frac{\delta(\sqrt{-g}L_{m,eff})}{\delta g^{\mu\nu}}\\ \nonumber
&&=T_{\mu\nu}+\frac{
\alpha}{\kappa}\left(\frac{1}{2}
T_{\alpha\beta}T^{\alpha\beta}g_{\mu\nu}+2L_{m}(T_{\mu \nu }-\frac{1}{2}g_{\mu \nu }T)+TT_{\mu
\nu }-2T_{\mu }^{\alpha }T_{\nu \alpha }+4T^{\alpha \beta }\frac{\partial
^{2}L_{m}}{\partial g^{\mu \nu }\partial g^{\alpha \beta }}
\right)~\label{teff}~.
%&&T_{\mu\nu}=-\frac{2}{\sqrt{-g}}\frac{\delta(\sqrt{-g}L_m)}{\delta g^{\mu\nu}}=
%g_{\mu \nu }L_{m}-2\frac{\partial L_{m}}{\partial g^{\mu \nu }}\label{teff}~.
\end{eqnarray} which clearly indicate that further degrees of freedom related to EMSG can be formally dealt under the standard of perfect fluids. See for example \cite{Capozziello:2018ddp}. 
The above equations allow us to show that $\nabla^\mu T^{\text{eff}}_{\mu\nu}=0$ not $\nabla^\mu T_{\mu\nu}=0$ \footnote{
Since EMT is a delicate issue of GR, as well as of any modified  theory of gravity,  it is worthy of further
discussion. First of all, it is necessary to point out  that the equation $\nabla^\mu T_{\mu\nu}=0$, in  curved spacetime, does not has
the same interpretation of conservation law  as $\partial^\mu T_{\mu\nu}=0$ in  absence of gravity. Actually, equation
$\nabla^\mu T_{\mu\nu}=0$ is a consequence of the contracted Bianchi Identity $\nabla^\mu G_{\mu_\nu}=0$ and expresses not only the conservation
of energy-momentum tensor but also the exchange of energy and momentum of matter with gravitational field \cite{Book:Padmanabah}. Thanks to the Equivalence Principle,  one can locally adopt an \emph{Inertial Frame} and then   $\nabla^\mu T_{\mu\nu}=0$ can be interpreted as a \emph{local conservation law}.
As a result, the meaning of energy conservation,  in the context of GR and modified gravities as well,
is  a {\it local concept} and, in principle, it is not a general conserved quantity in  presence of farther degrees of freedom and fields. Generally speaking,  energy localization is one of the challenging issues
in GR. Its investigation, in the context of alternative theories of gravity,  attracted a lot of attention in recent years leading to generalizations of  the gravitational stress-energy tensor, i.e. the so-called energy-momentum complex \cite{Capozziello:2017xla,Capozziello:2018qcp,Abedi:2018eyz}. However, it is worth stressing that
any consistent theory of gravity, in particular any  extension of GR, must respect the Bianchi Identity and, subsequently, the local conservation equation. However,  due to the appearance of corrections in modified theories of gravity (e.g. higher-order curvature terms, further scalar fields  or $(T^{\mu\nu}T_{\mu\nu})^n$ as in the present case), we can deal with  generalized local conservation laws as $\nabla^\mu T^{\text{eff}}_{\mu\nu}=0$. In other words, these  corrections can be dealt as  effective fluids whose  interpretation differs from the conventional matter fluids
commonly considered as  sources of the Einstein field equations \cite{Capozziello:2013vna,Capozziello:2014bqa}.
In the  context of modified gravity, such as EMSG  considered here,  imposing the  divergence-free condition in the r.h.s.  of \eqref{Ef},  the equation $\nabla^\mu T_{\mu\nu}=0$ unlike the common imagine \cite{Katirci:2013okf} is no longer a criterion to measure the healthiness of theory but, instead, its generalized counterpart $\nabla^\mu T^{\text{eff}}_{\mu\nu}=0$ has to be considered.
However, there is a noteworthy point related to the generalization of energy conditions in modified gravity. See \cite{Capozziello:2013vna,Capozziello:2014bqa} for a detailed  discussion. In this regard, it is helpful to refer  also to \cite{Koivisto:2005yk} where  conservation conditions are discussed for a wide class of models with matter non-minimally coupled to gravity. There   extended conservation laws are obtained. In particular, the conservation issue can be probed from the perspective of particle creation. In \cite{Harko:2015pma},  non-conservation issue related to  non-minimal coupling between the matter  and curvature is discussed. These results are   criticized in \cite{Azevedo:2019krx}. The key argument of the criticism  in \cite{Azevedo:2019krx} is based on the claim that  non-minimal coupling, in essence, induces a change in the particle-momentum at a cosmological timescale, which is not relevant for  the particle creation process. So, in the framework of particle creation phenomena, the EMSG as a theory with non-minimally coupling between   curvature and matter, still can preserve the conservation properties.}.
It is assumed here that $L_m$ is free of derivatives of the metric
components.
By considering the perfect fluid form for the EMT, we have
\begin{eqnarray}\label{pf}
T_{\mu \nu }=(\rho +p)U_{\mu }U_{\nu }+pg_{\mu \nu }, ~~~~~T_{\mu\nu}T^{\mu\nu}=\rho^2+3p^2
\end{eqnarray}
where $\rho $, $p$ and $U_{\mu}$ respectively denote the energy density, the pressure and the co-moving four-velocity satisfying the conditions, $U_{\mu} U^{\mu}= -1$, $\nabla_{\nu}U^{\mu}U_{\mu}= 0$. Now by taking the Lagrangian $L_m=p$\footnote{As it is well-known $L_m=p$ is not the unique choice for the matter Lagrangian representing the EMT perfect fluid.  In fact, one can consider  other cases like,  for instance, $L_m=-\rho$. According to some  results \cite{Schutz:1970my,Brown:1992kc} this is not problematic in the context of GR, but in some modified gravity models, such as EMSG, it is expected that the choice of  $L_m$ affects the dynamics, see \cite{Bertolami:2008ab,Faraoni:2009rk}. However, it is worth noticing that, in case of minimal coupling of the fluid with  gravity, the two matter Lagrangian $L_m=p$ and $L_m=-\rho$ are equivalent.	While there are no definitive criterion  for the choice   of  these two matter Lagrangians,  the former seems more natural. Hence, we choose it  to conform with the EMSG literature \cite{Barbar:2019rfn}.}, the last term in the right hand side of Eq.\eqref{tef} cancels so that after inserting Eq.\eqref{pf} into it, the modified Einstein’s field equations finally read as 
\begin{equation}\label{fieldeq}
\begin{split}
R_{\mu\nu}-\frac{1}{2}Rg_{\mu\nu}+\Lambda g_{\mu\nu}=\kappa ((\rho & +p)U_{\mu }U_{\nu
}+pg_{\mu \nu }) +\alpha \left(\frac{1}{2}(\rho ^{2}+3p^{2})g_{\mu
\nu }+(\rho +p)(\rho +3p)U_{\mu }U_{\nu }\right).
\end{split}
\end{equation}
The right-hand side of the above equation indicates that we no longer deal with a standard perfect fluid but with an effective fluid.
More precisely,  the whole budget of EMT does not come from standard matter fields only  but there
are  other contributions coming from the non-Einsteinian part of the gravitational interaction \cite{Carloni:2004kp,Carloni:2007eu,SantosdaCosta:2018ovq}.
The cosmological applications of such effective fluids, arising from generalized theories of gravity,
have been considered in recent years, see e.g. \cite{Capozziello:2018ddp,Capozziello:2019wfi,Capozziello:2019qlt}. The paradigm is that  contributions coming from geometric degrees of freedom or further fields are brought to the r.h.s. of field equations and then  the r.h.s. is globally considered as a source satisfying the Bianchi Identity.
We want to investigate the implications of these field equations in cosmology. Therefore, we consider the FLRW cosmological models. Using the FLRW metric
\begin{equation}\label{metric}
ds^{2}=-dt^{2}+a^{2}(t)\left( \frac{dr^{2}}{1-kr^{2}}+r^{2}\left( d\theta
^{2}+\sin ^{2}\theta d\phi ^{2}\right) \right) ,
\end{equation}
characterized by the expansion scale factor $a(t)$ and the  constant curvature parameter $k=-1,0,1$
(corresponding to spatially open, flat and closed universes respectively),  we obtain the following
modified dynamical equations
\begin{equation}\label{ttfriedmann}
\frac{\dot{a}^2}{a^2}=-\frac{k}{a^{2}}+\frac{\Lambda }{3}
+\kappa \frac{\rho }{3}+\frac{\alpha }{6}\left(\rho ^{2}+3p^{2}+8\rho p\right) ,
\end{equation}
and
\begin{equation}\label{frttaccel}
\frac{\ddot{a}}{a}=-\kappa \frac{\rho +3p}{6}+\frac{\Lambda }{3}-\frac{\alpha
}{3}\left( \rho ^{2}+3p^{2}+2\rho p\right) ,
\end{equation} for an isotropic and homogeneous universe where dot means derivative with respect to the cosmic time. Assuming that  in a spacetime with the FLRW background metric, the matter field obeys
a barotropic equation of state (EoS), $p/\rho=w=Const $, then we can
 extend the dynamical equations expressed in the following final forms
\begin{equation}\label{frttfriedmannb}
\frac{\dot{a}^2}{a^2}=-\frac{k}{a^{2}}+\frac{\Lambda }{3}
+\kappa \frac{\rho }{3}+\frac{\alpha A_1(w)}{3}\rho ^{2},
\end{equation}
and
\begin{equation}\label{frttaccelb}
\frac{\ddot{a}}{a}=-\kappa \frac{1+3w}{6}\rho +\frac{\Lambda }{3}-\frac{\alpha A_2(w)
}{3}\rho ^{2},
\end{equation}
with constants $A_{1,2}(w)$ designated by
\begin{eqnarray}
A_1(w)\equiv \frac{3}{2}w^{2}+4w+\frac{1}{2},\quad
A_2(w)\equiv 3w^2+2w+1~.
\end{eqnarray}
Merging the above dynamical equations, we  obtain
\begin{equation}\label{frttaccelc}
\dot{H}=-\frac{1+w}{2}\rho-\frac{\alpha}{3}(A_1+A_2)\rho^2+\frac{k}{a^2},
\end{equation} where $H=\frac{\dot{a}}{a}$ and $\dot{H}=\frac{\ddot{a}}{a}
-H^2$.
Differentiating the Friedman equation \eqref{frttfriedmannb}, we see that
the GR-based conservation equation also is modified as
\begin{equation}\label{frttcontb}
\dot{\rho}+3\frac{\dot{a}}{a} (1+w)\rho
\left(\frac{\kappa+\alpha\rho (1+3w)}{\kappa+2\alpha\rho A_1(w)}\right)=0.
\end{equation}
The above modified conservation equation is the cosmological counterpart of $\nabla^\mu T^{\text{eff}}_{\mu\nu}=0$. It means that we do not have to demand conservation  of the ordinary $T_{\mu\nu}$, but the whole right-hand side of Einstein equation \eqref{Ef}, that is conservation  of $T^{\text{eff}}_{\mu\nu}$.
As a consistency check, one can see immediately that if $\alpha=0$, then dynamical equations
\eqref{frttfriedmannb}, \eqref{frttaccelb} as well as the modified conservation equation \eqref{frttcontb},
 reproduce exactly  GR behaviors. Due to the fact that the source of the modification in the
gravity model under consideration comes from the extension of EMT, it is quite natural that
the modified terms in dynamical equations \eqref{frttfriedmannb} and \eqref{frttaccelb} appear as
$\mathcal{O}(\rho^{2})$. Eqs. \eqref{frttfriedmannb} and \eqref{frttaccelb}, together with
\eqref{frttaccelc}, form a complete dynamical set to investigate the existence and
stability status of static solutions, as a prerequisite for the realization of EU. Hereinafter, for simplicity
we will set $\kappa=1$ in our calculations. However, the other values of $\kappa$ can be considered under the same standard.

%%%%%%%%%%%%%%%%%%%%%%%%%%%%%%%%%%%%%%%%%%%%%%%%%%%%%%%%%%%%%%%%%%%
\section{first order cosmological dynamics and stability analysis of
Einstein static solutions}\label{ESU}

The system of Eqs. \eqref{frttfriedmannb}-\eqref{frttaccelc} admit static solutions characterized by $\dot{a}=\ddot{a}=\dot{\rho}=0$
(i.e. $a_{ES}=\rho_{ES}=$const) which address the ESU phase.
Let us start with a flat FLRW metric. Considering the cosmological dynamical system with
$k=0$ and applying this condition in Eqs. \eqref{frttfriedmannb} and \eqref{frttaccelb}, we obtain the
critical points (CPs) as follows
\begin{equation}\label{cp}
\mbox{CP:~A}~~~\rho_{ES}=\frac{-1\pm\sqrt{1-4\Lambda A_1\alpha}}{2A_1\alpha},
\end{equation} with any positive arbitrary value of $\frac{1}{a_{ES}^2}$. However, for the above CPs to be
physically meaningful, we have to check the non-negative energy density condition $\rho_{ES}\geq0$.
Thus, the CPs $A_{\pm}$, are physically meaningful if
\begin{eqnarray}\label{cpap}
\begin{array}{ll}
\Lambda=0,~~\alpha\neq0,~~\bigg(w<-\frac{4+ \sqrt{13}}{3},~~\mbox{or}~~-\frac{4+ \sqrt{13}}{3} < w < \frac{-4 + \sqrt{13}}{3}~~\mbox{or}~~w>\frac{-4 + \sqrt{13}}{3}\bigg) \\
\Lambda<0,~~\alpha<0,~~\bigg(-\frac{4+ \sqrt{13}}{3} < w < \frac{-4 + \sqrt{13}}{3}~~ \mbox{or}~~\frac{-4 + \sqrt{13}}{3} < w \leq-\frac{4}{3}+
\frac{\sqrt{26+\frac{3}{\alpha\Lambda}}}{3\sqrt{2}} \bigg) \\
\Lambda<0,~~0<\alpha<-\frac{3}{26\Lambda}, ~~\bigg(-\frac{4+ \sqrt{13}}{3} < w < \frac{-4 + \sqrt{13}}{3},~~\mbox{or}~~w>\frac{-4 + \sqrt{13}}{3} \bigg)\\
\Lambda<0,~~\alpha>-\frac{3}{26\Lambda},~~ \bigg(-\frac{4+ \sqrt{13}}{3} < w \leq-\frac{4}{3}-
\frac{\sqrt{26+\frac{3}{\alpha\Lambda}}}{3\sqrt{2}},~~ \mbox{or}~~ -\frac{4}{3}+
\frac{\sqrt{26+\frac{3}{\alpha\Lambda}}}{3\sqrt{2}}\leq w < \frac{-4 + \sqrt{13}}{3},~~\mbox{or}~~w>\frac{-4 + \sqrt{13}}{3}\bigg) \\
\end{array}
\end{eqnarray}
and
\begin{eqnarray}\label{cpan}
\begin{array}{ll}
\Lambda\geq0,~~\alpha<0,~~w > \frac{-4 + \sqrt{13}}{3}   \\
\Lambda\geq0,~~\alpha>0,~~-\frac{4+ \sqrt{13}}{3} < w < \frac{-4 + \sqrt{13}}{3}  \\
\Lambda<0,~~0<\alpha<-\frac{3}{26\Lambda}, ~~-\frac{4+ \sqrt{13}}{3} < w < \frac{-4 + \sqrt{13}}{3}, \\
\Lambda<0,~~\alpha>-\frac{3}{26\Lambda},~~\bigg( -\frac{4+ \sqrt{13}}{3} < w \leq-\frac{4}{3}-
\frac{\sqrt{26+\frac{3}{\alpha\Lambda}}}{3\sqrt{2}},~~ \mbox{or}~~ -\frac{4}{3}+
\frac{\sqrt{26+\frac{3}{\alpha\Lambda}}}{3\sqrt{2}}\leq w < \frac{-4 + \sqrt{13}}{3},~~\mbox{or}~~w>\frac{-4 + \sqrt{13}}{3}\bigg) \\
\end{array}
\end{eqnarray} respectively. So in case of satisfying any of conditions released in Eqs.
\eqref{cpap} and \eqref{cpan}, the physical meaning of the CPs $A_{\pm}$ can be guaranteed, respectively.

The  dynamical systems with $k=\pm1$  can be obtained by applying the condition related to ESU
in Eqs. \eqref{frttfriedmannb} and \eqref{frttaccelb}. We obtain the following CPs
\begin{eqnarray}\label{cpcd}
\begin{array}{ll}
\mbox{CP:~C}~~~\frac{1}{a_{ES}^2}=\frac{(1+3w)A_1\big(-1\pm\sqrt{1-\frac{16\Lambda A_2\alpha}{(1+3w)^2}}\big)^2}{48\alpha A_2^2}+
\frac{(1+3w)\big(-1\pm\sqrt{1-\frac{16\Lambda A_2\alpha}{(1+3w)^2}}\big)}{12\alpha A_2}+\frac{\Lambda}{3},\\
\mbox{CP:~D}~~~\frac{1}{a_{ES}^2}=
-\frac{(1+3w)A_1\big(-1\pm\sqrt{1-\frac{16\Lambda A_2\alpha}{(1+3w)^2}}\big)^2}{48\alpha A_2^2}-
\frac{(1+3w)\big(-1\pm\sqrt{1-\frac{16\Lambda \alpha A_2}{(1+3w)^2}}\big)}{12\alpha A_2}-\frac{\Lambda}{3},
\end{array}
\end{eqnarray}
with the energy density
\begin{eqnarray}
\rho_{ES}=
\frac{(1+3w)\big(-1\pm\sqrt{1-\frac{16\Lambda A_2\alpha}{(1+3w)^2}}\big)}{4A_2\alpha}
\end{eqnarray}
for both CPs.
Note that the CPs $C_{\pm}$ and $D_{\pm}$ come from the closed
and open universes, respectively. Here because of the added conditions $\frac{1}{a_{ES}^2}>0$ and
$\rho_{ES}\geq0$, the extraction of the limits on   parameters $(w,~\alpha,~\Lambda)$ to ensure
the physical meaning of the aforementioned CPs is very complicated. However, we will do that for CPs $C_{\pm}$
and $D_{\pm}$, via the parameter space plots in terms of $w-\alpha$ with some positive, negative and zero fixed
values of $\Lambda$. For similar analysis in extended gravity theories, see e.g. \cite{Carloni:2004kp,Carloni:2007eu,SantosdaCosta:2018ovq}.
\begin{figure}[!ht]
	\begin{center}
			\includegraphics[scale=0.45]{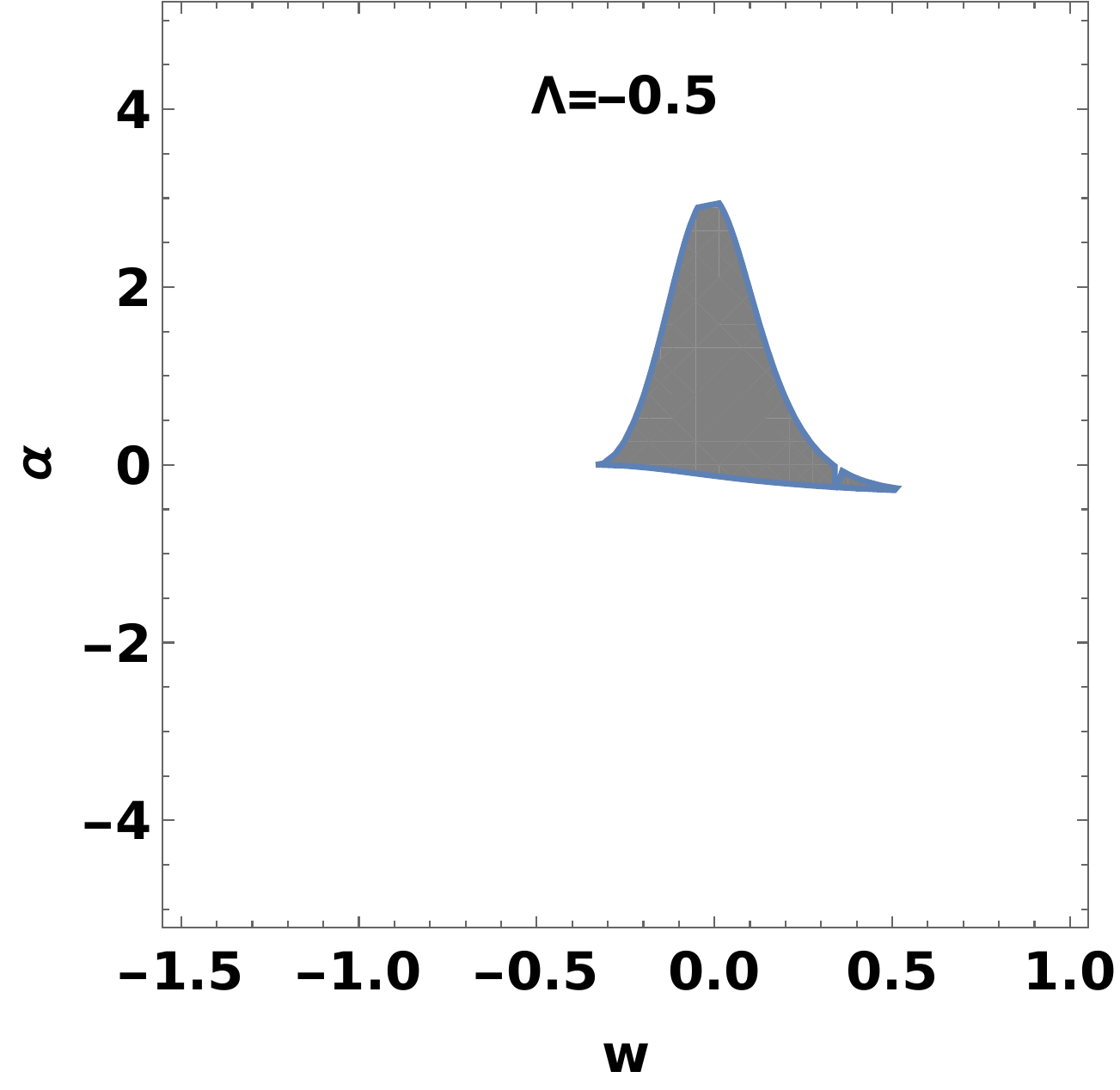}
				\includegraphics[scale=0.45]{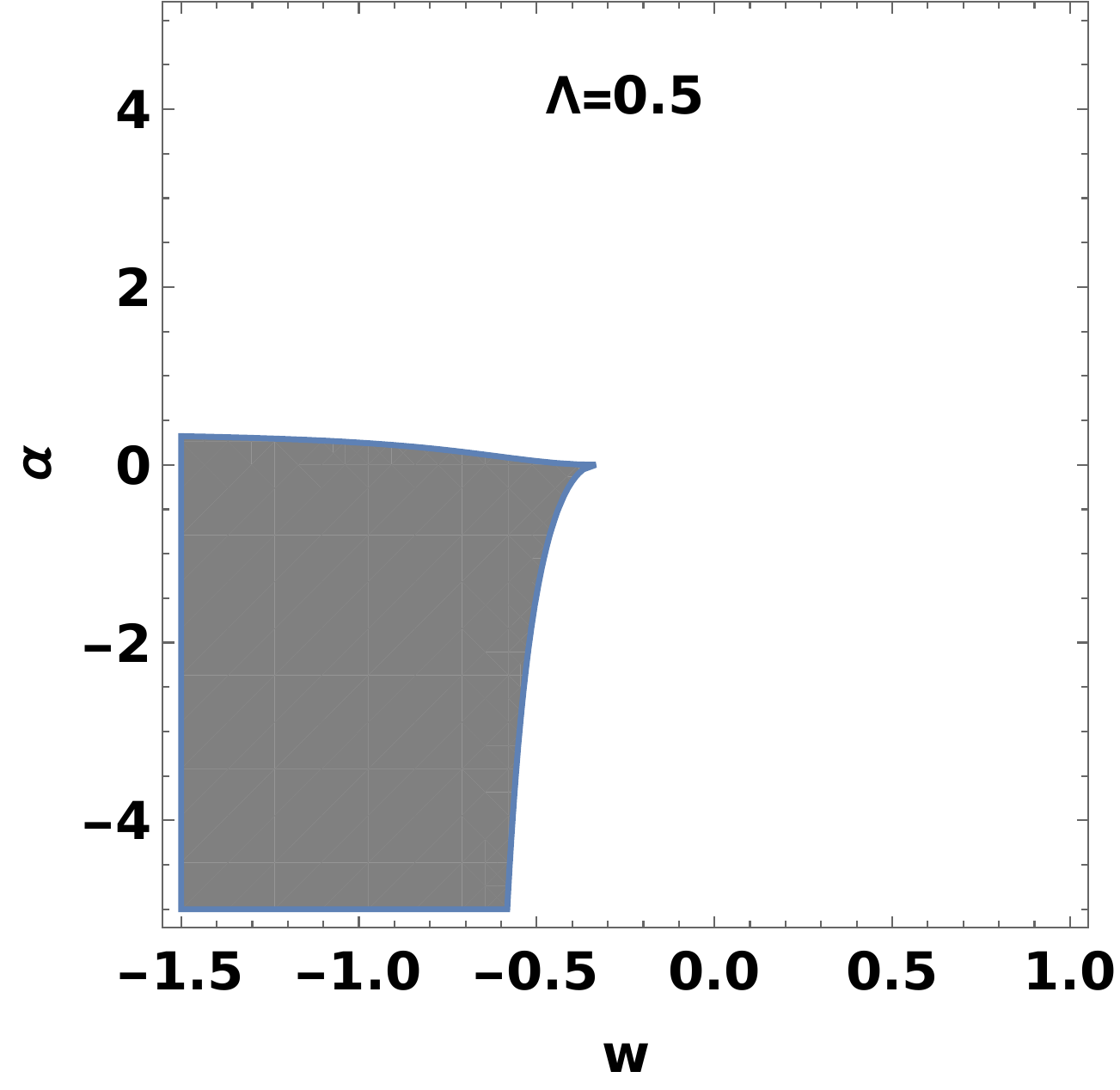}\\
                 \includegraphics[scale=0.45]{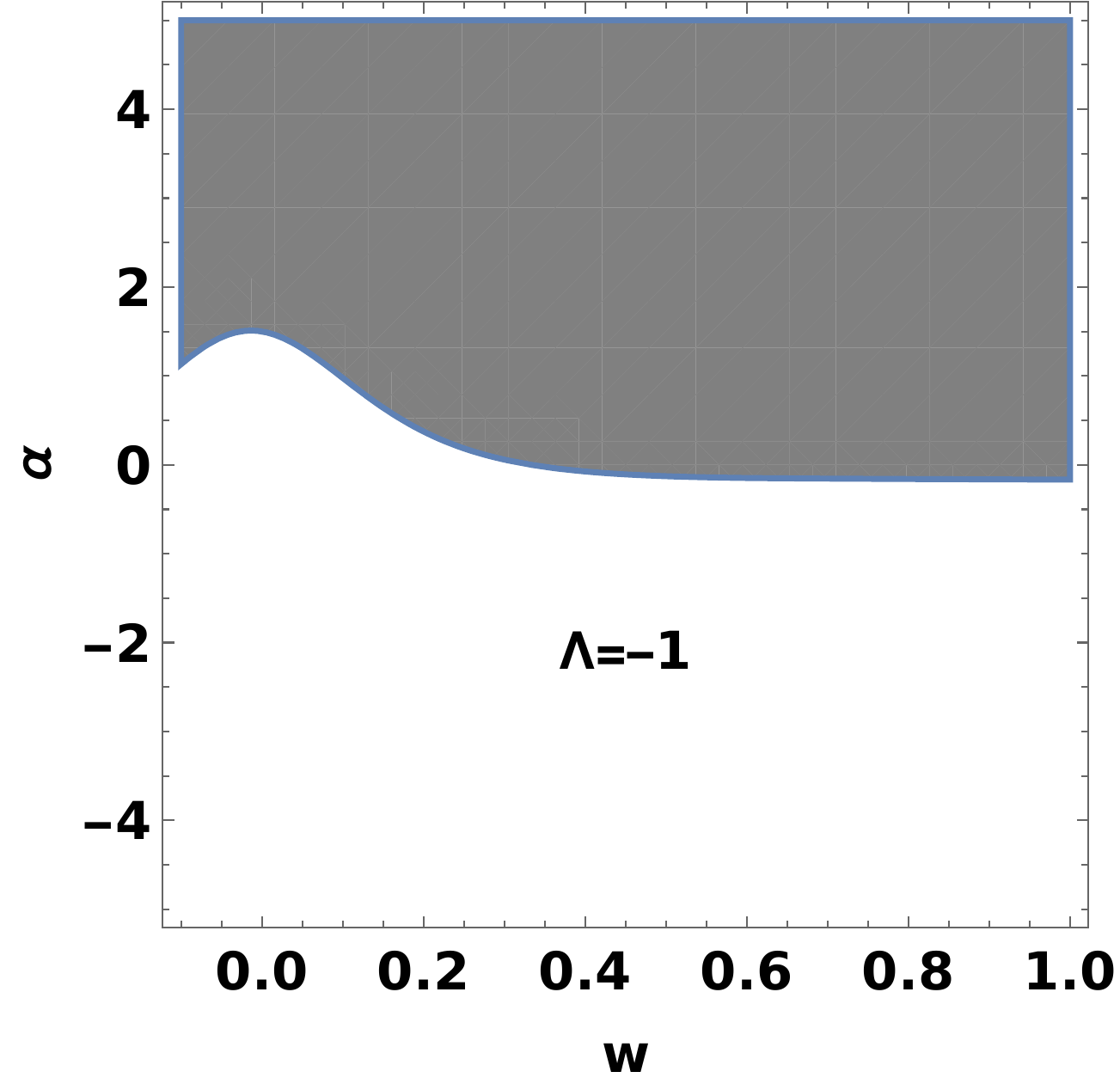}
				\includegraphics[scale=0.45]{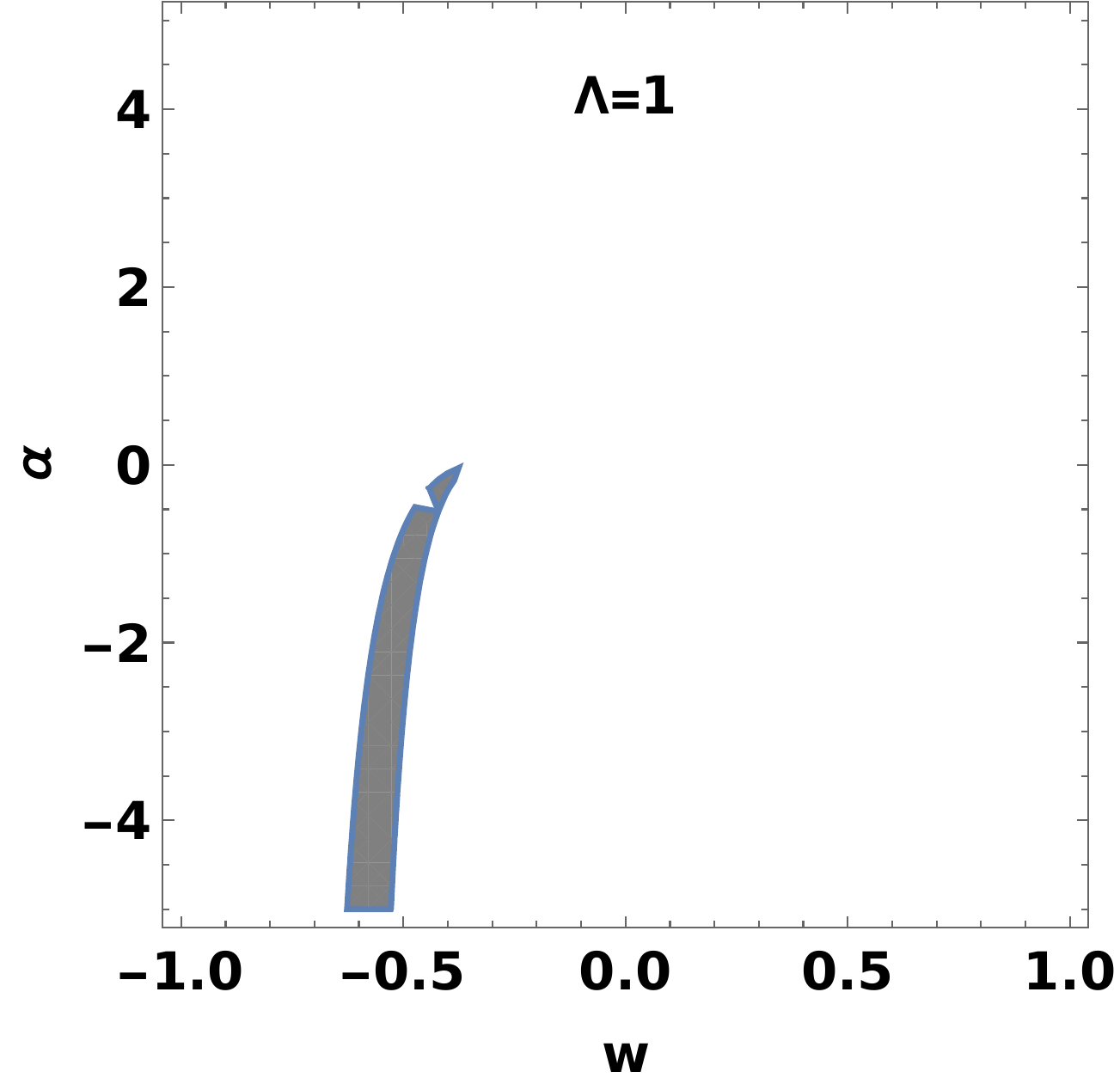}
		\caption{Regions of existence in the $(w,\alpha)$ parameter space for the stable CPs $C_+$ (up row) and $D_+$
(bottom row). $C_+$ and $D_+$ refer to the FLRW with $k=1$ and $k=-1$, respectively. }
		\label{cdp}
	\end{center}
\end{figure}
\begin{figure}[!ht]
	\begin{center}
			\includegraphics[scale=0.4]{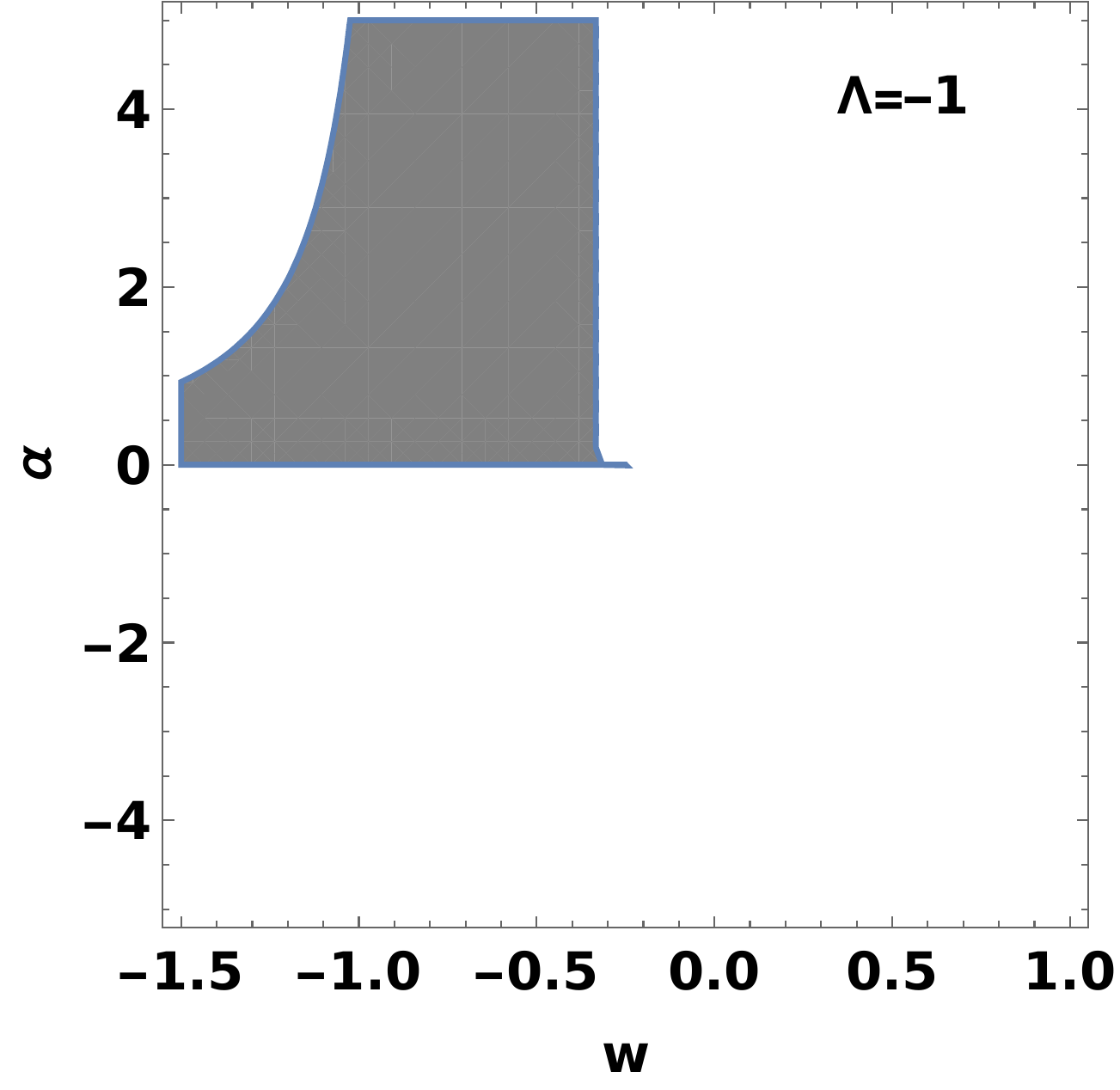}
			\includegraphics[scale=0.4]{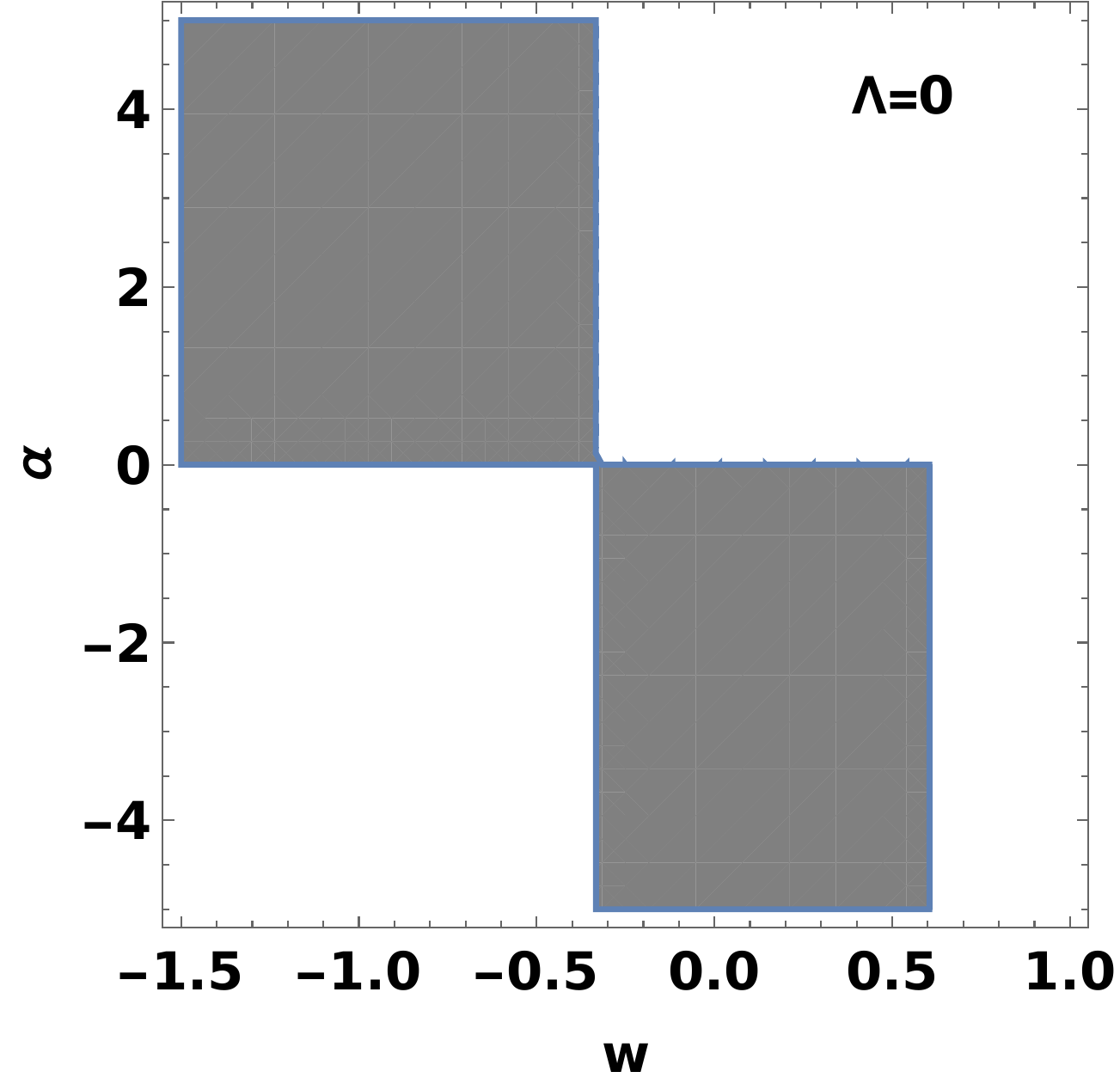}
            \includegraphics[scale=0.4]{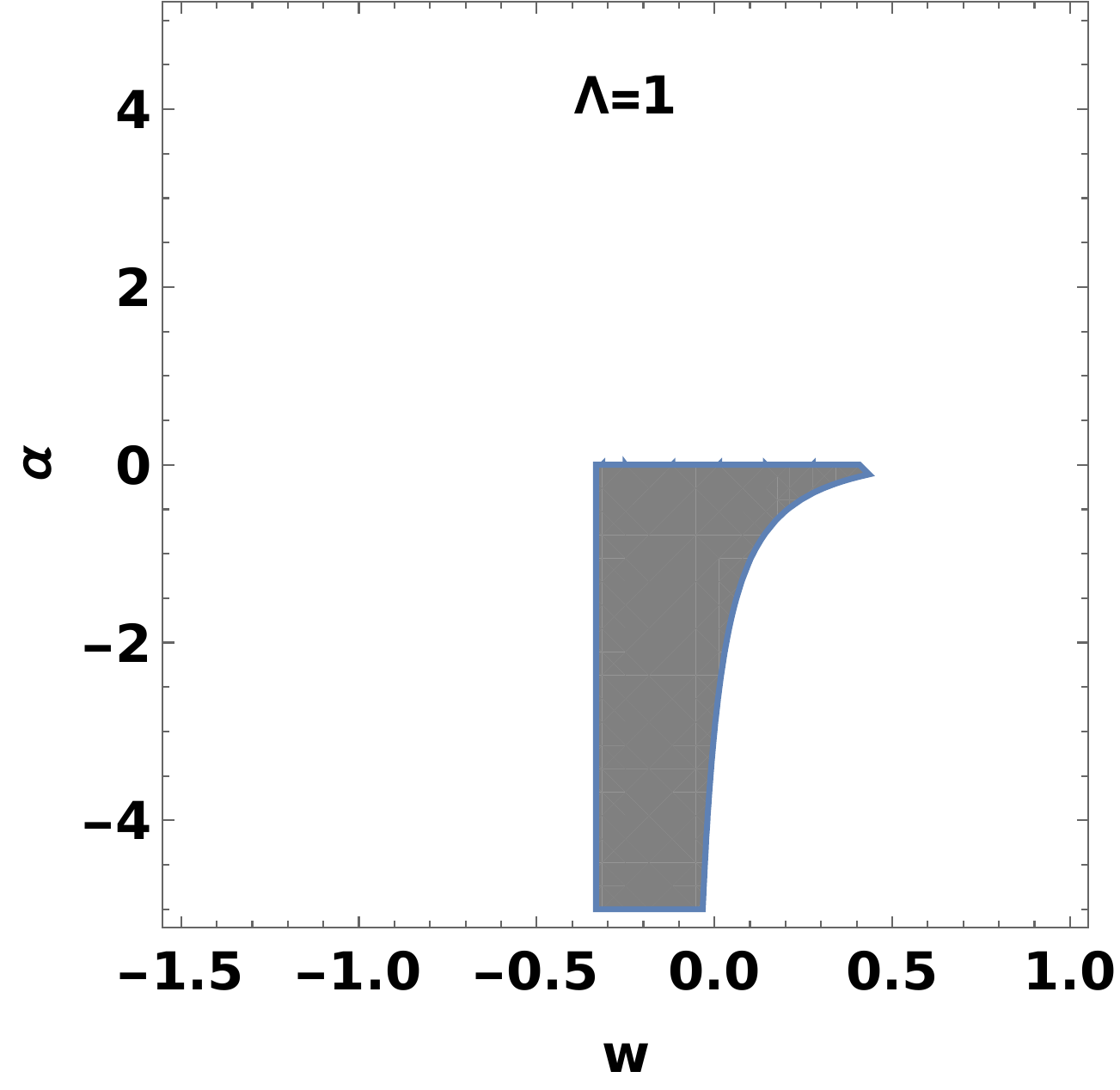}\\
            \includegraphics[scale=0.4]{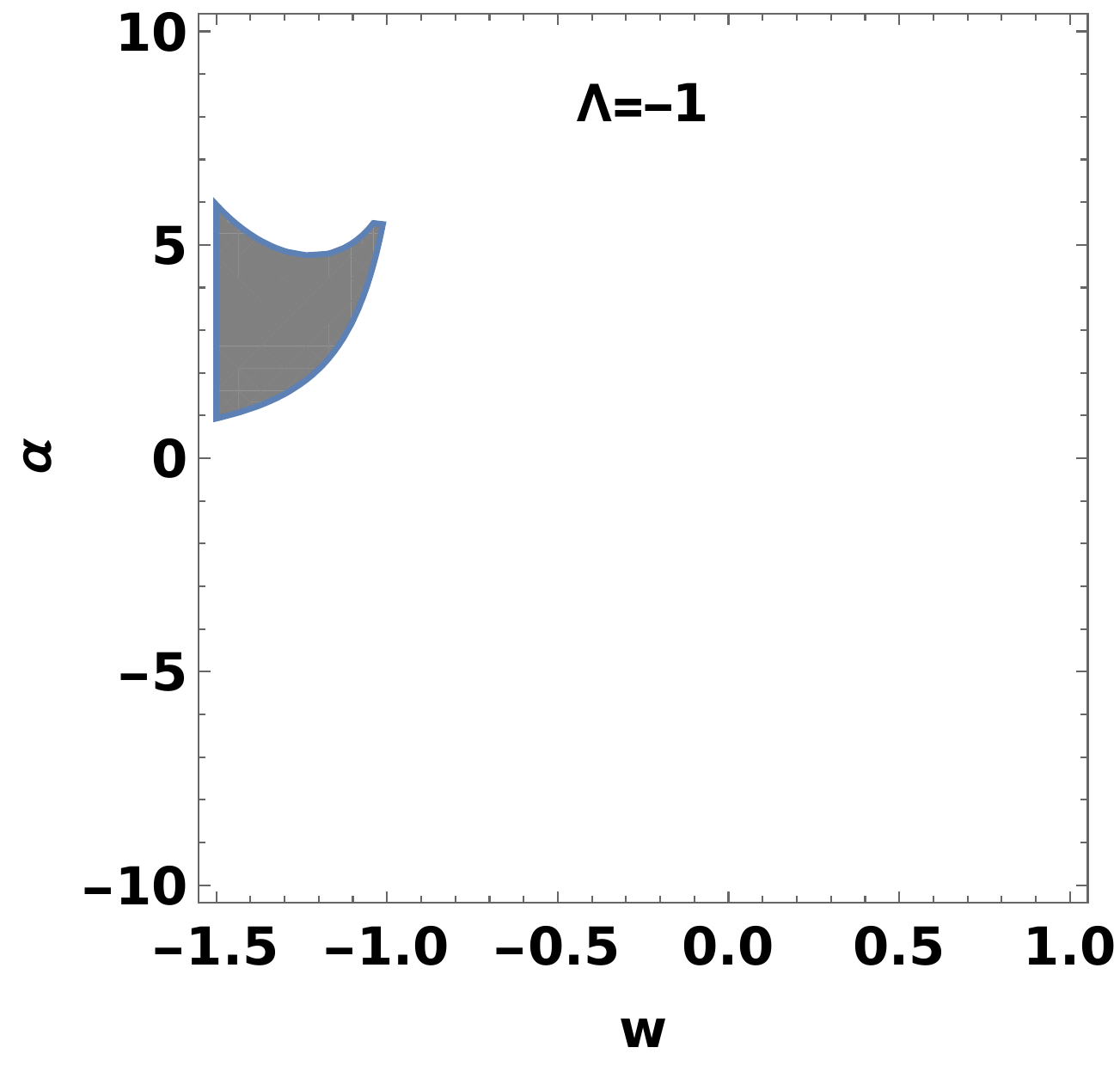}
            \includegraphics[scale=0.4]{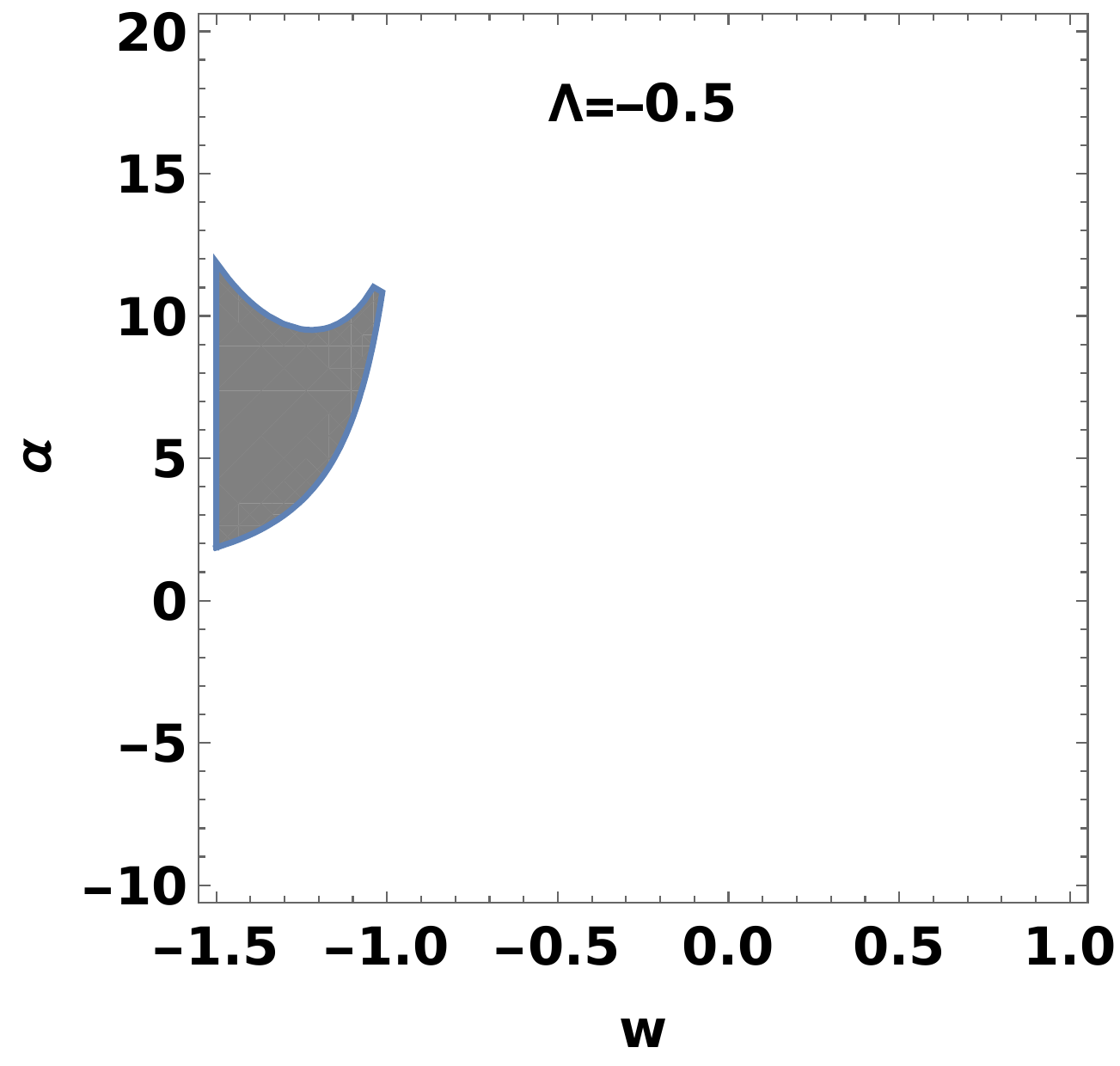}
            \includegraphics[scale=0.4]{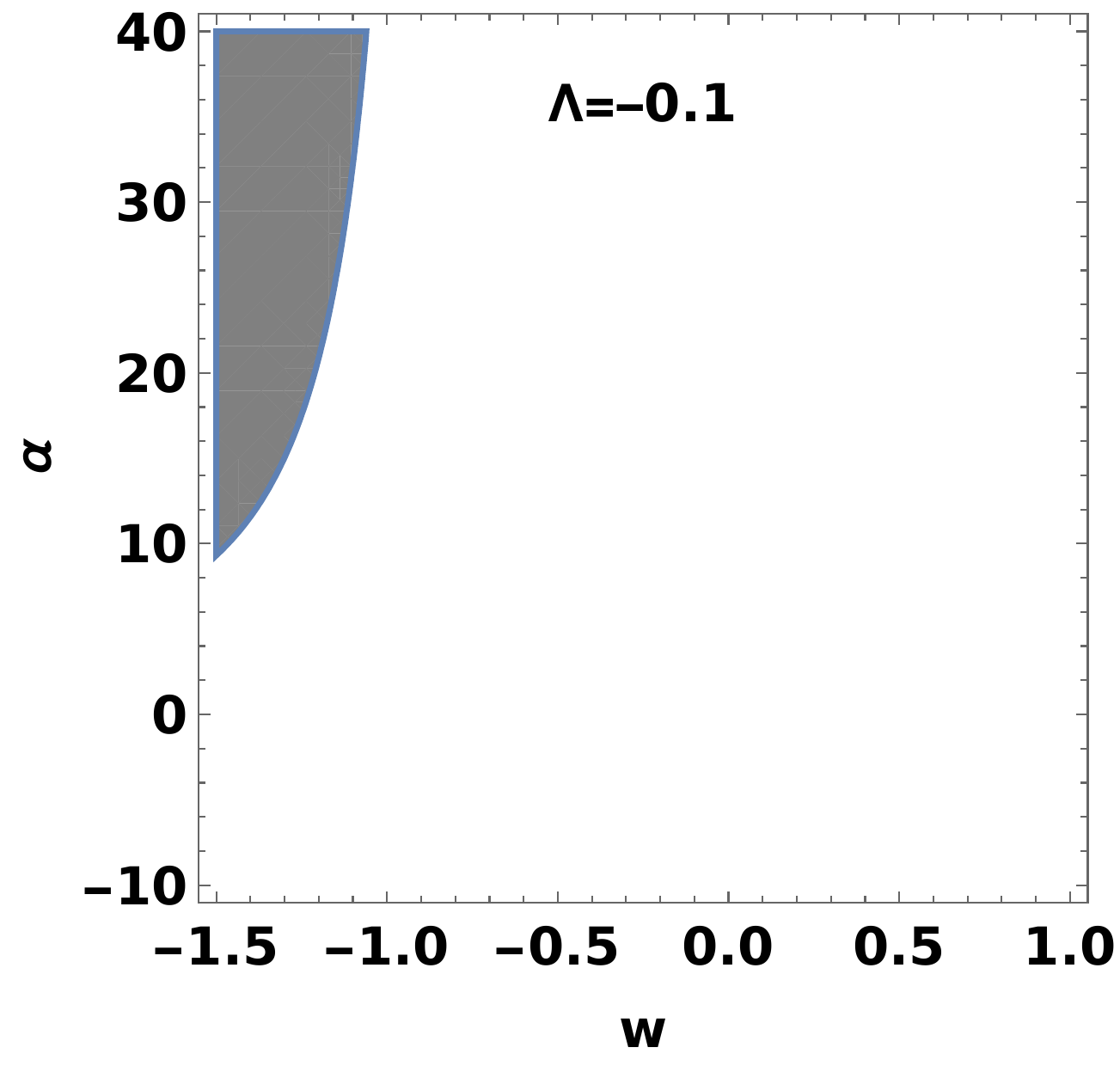}
		\caption{Regions of existence in the $(w,\alpha)$ parameter space for the stable CPs $C_-$ (up row) and $D_-$
(bottom row). $C_-$ and $D_-$ refer to the FLRW with $k=1$ and $k=-1$, respectively. }
		\label{cdn}
	\end{center}
\end{figure}

\begin{figure}[!ht]
	\begin{center}
			\includegraphics[scale=0.4]{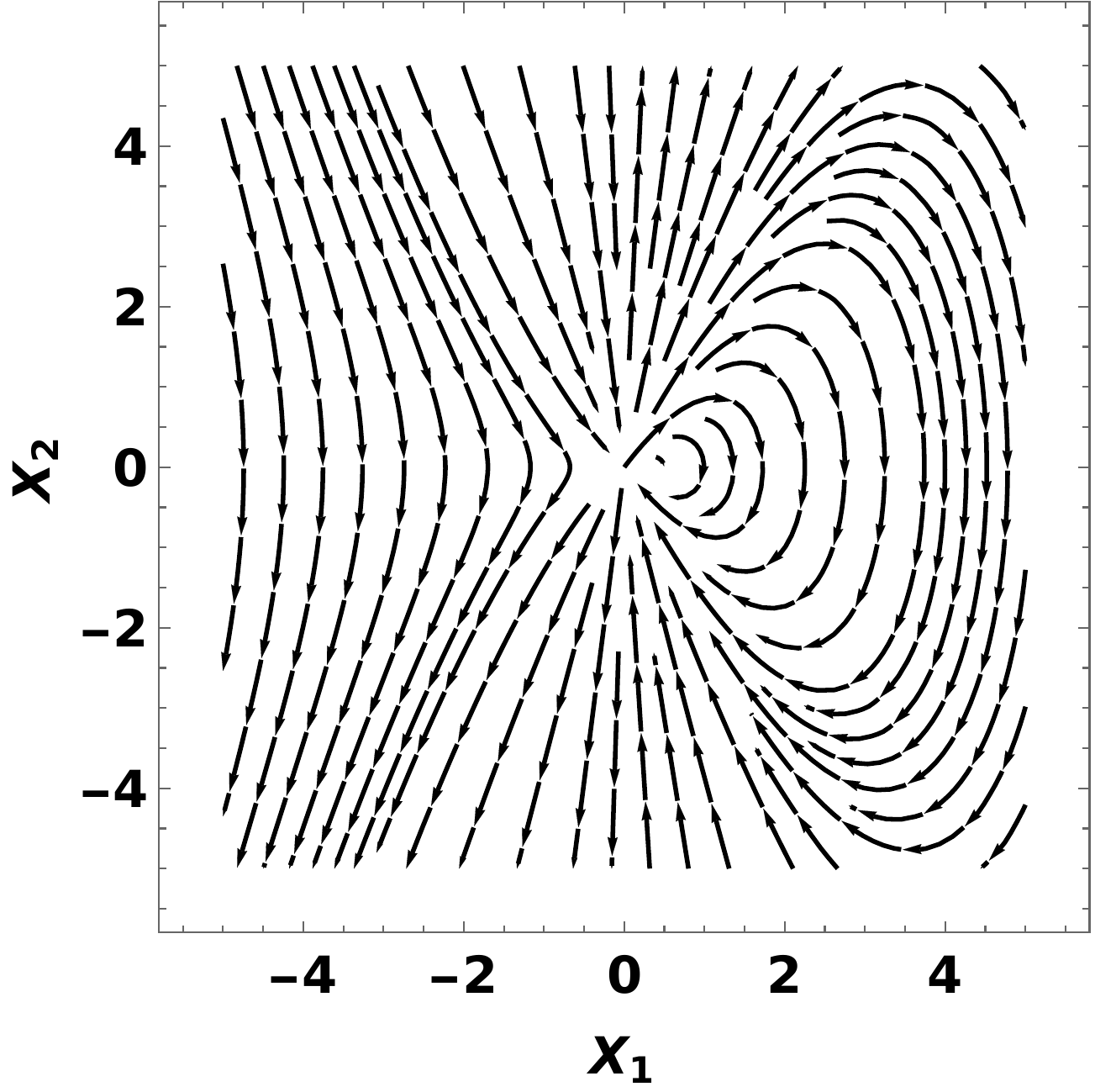}
			\includegraphics[scale=0.4]{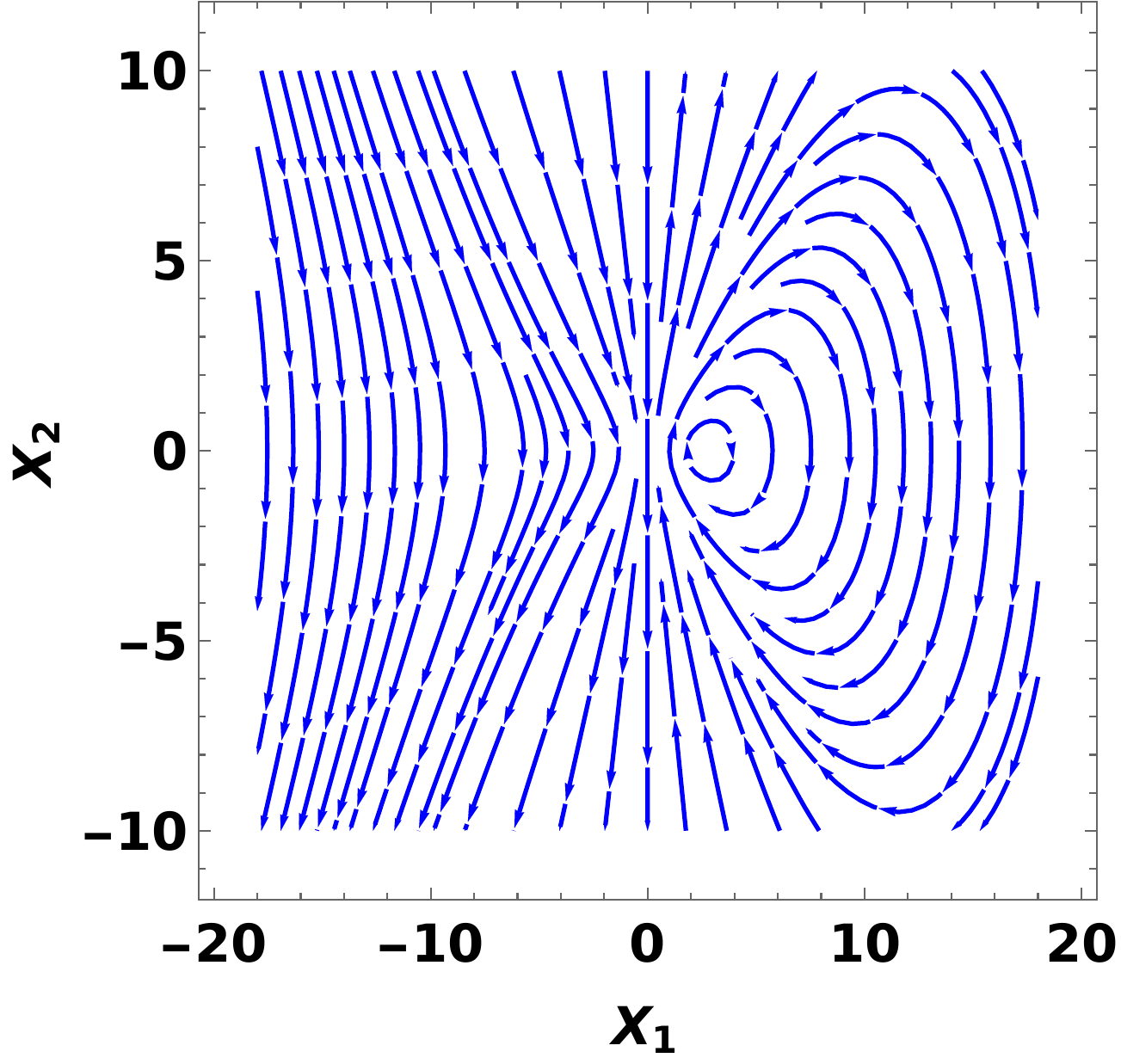}
            \includegraphics[scale=0.4]{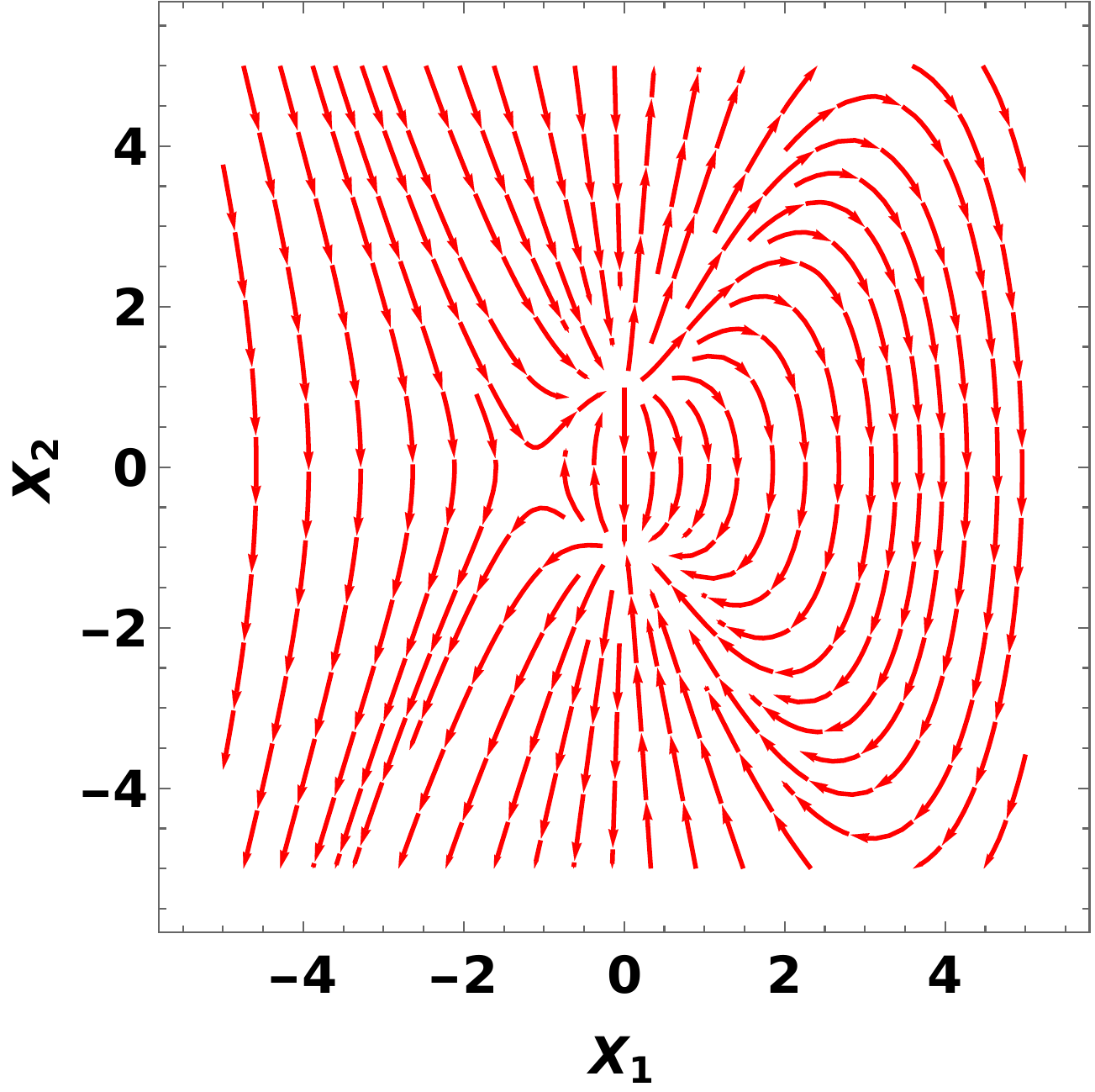}\\
            \caption{The phase diagram in $(a, \dot{a})$ or $(X_1, X_2)$ space for
            spatially flat, closed and open cosmology with the CPs $A_-,~C_+$ and $D_+$, respectively from left to right panels.
            We use numerical values $(0,2,0)$, $(-0.8,-4,0.5)$ and $(0,5,-1)$ for the set parameters $(w,\alpha,\Lambda)$, respectively. }
		\label{pd}
	\end{center}
\end{figure}

Now, in order  to investigate the stability of these CPs, we follow the strategy that the second order dynamics of the cosmological
model after introducing two variables $X_1=a,~~~X_2=\dot{a}$ can be reduced to the first order dynamics, as proposed for instance
in Refs. \cite{Carneiro:2009et, Wu:2009ah, Zhang:2013ykz, Huang:2015zma}.
Subsequently, we have a system of  coupled equations as follows
\begin{eqnarray}\label{Dy}
\begin{array}{ll}
\dot{X}_1=X_2=\mathcal{G}_1(X_1,X_2)~,\\
\dot{X}_2=\frac{X_2^2}{X_1}-\bigg(\frac{1+w}{2}\rho+\frac{\alpha}{3}(A_1+A_2)\rho^2\bigg)X_1+\frac{k}{X_1^2}=
\mathcal{G}_2(X_1,X_2)~.
\end{array}
\end{eqnarray}
The behavior of the coupled equations near the critical points can be studied by the  Jacobian matrix of the coupled system. More precisely, the eigenvalues of the Jacobian matrix are related to the stability or instability of the system. Moreover,
we extract the eigenvalues squared, $\lambda^2$, via the following Jacobian matrix
\begin{eqnarray}\label{e3-6}
J\bigg(\mathcal{G}_1(X_1,X_2),\mathcal{G}_2(X_1,X_2)\bigg)=\left(
 \begin{array}{cc}
 \frac{\partial \mathcal{G}_1}{\partial X_1} & \frac{\partial \mathcal{G}_1}{\partial X_2}\\
 \frac{\partial \mathcal{G}_2}{\partial X_1} & \frac{\partial \mathcal{G}_2}{\partial X_2}\\
 \end{array}
  \right)~~\Rightarrow~~~\lambda^2=\frac{\partial \mathcal{G}_2}{\partial X_1}|_{(a_{ES},\rho_{ES})}= \nonumber \\
  -\bigg(\frac{3}{2}\alpha w^2+2\alpha w+\frac{\alpha}{2}\bigg)\rho_{ES}^2-\frac{1+w}{2}\rho_{ES}-\frac{k}{a_{ES}^2}~.
  \end{eqnarray}
 When $\lambda^2<0$, we can consider the
relevant physical CP as a stable ESU. Physically, $\lambda^2<0$
means that the relevant CP is a stable concentric center and if it undergoes some small perturbations,
it does not collapse. In contrast, when $\lambda^2>0$, we expect to encounter an unstable saddle point
\footnote{It is important to mention that  $\lambda^2>0$  does not always address an unstable saddle point since
 conditions $\frac{\partial \mathcal{G}_1}{\partial X_1} \neq\frac{\partial \mathcal{G}_2}{\partial X_2}$
and $\frac{\partial \mathcal{G}_1}{\partial X_2}=0=\frac{\partial \mathcal{G}_2}{\partial X_2}$, in the Jacobian matrix (\ref{e3-6})
can lead to $\lambda^2>0$ which represents a stable node.}.

Let us start our analysis with CPs $A_{\pm}$ which come from the
flat FLRW case ($k=0$). By imposing the conditions $\lambda^2<0$ as well as $\rho_{ES}\geq0$
simultaneously, we get the following constraints on the set $(w,\alpha,\Lambda)$
\begin{eqnarray}\label{con1}
\begin{array}{ll}
\Lambda<0,~~\alpha \leq \frac{3-\sqrt{13}}{4 \Lambda }~,~~-1<w<
   \sqrt{\frac{4}{54 \alpha \Lambda +9}}-\frac{1}{3}\\
 \Lambda<0,~~\frac{3-\sqrt{13}}{4 \Lambda }<\alpha \leq \frac{1}{2 \Lambda },~~
   \left(-1<w<\frac{\sqrt{13}-4}{3} ~\mbox{or}~ \frac{\sqrt{13}-4}{3}<w<
   \sqrt{\frac{4}{54 \alpha  \Lambda +9}}-\frac{1}{3} \right)\\
   \Lambda<0~,~~\frac{1}{2 \Lambda }<\alpha <0,~~ \left(-1<w<\frac{\sqrt{13}-4}{3},~\mbox{or} ~\frac{\sqrt{13}-4}{3}
   <w\leq \sqrt{\frac{26 \alpha  \Lambda +3}{18\alpha  \Lambda }}-\frac{4}{3}\right) \\
       \Lambda<0,~~0<\alpha \leq -\frac{1}{8 \Lambda },~~ \left(-1<w<\frac{\sqrt{13}-4}{3}~\mbox{or}~ w>\frac{\sqrt{13}-4}{3}
   \right) \\
\Lambda<0,~~   -\frac{1}{8 \Lambda }<\alpha \leq \frac{3}{4 \Lambda}+\frac{\sqrt{13}}{4 |\Lambda|},~~ \left(-\sqrt{\frac{4}{54 \alpha  \Lambda +9}}-\frac{1}{3}<w\leq -\sqrt{\frac{26 \alpha  \Lambda +3}{18\alpha  \Lambda }}-\frac{4}{3}~\mbox{or}~ \right. \\ \left. \sqrt{\frac{26 \alpha  \Lambda +3}{18\alpha  \Lambda }}-\frac{4}{3}\leq w<\frac{\sqrt{13}-4}{3}~\mbox{or}~
   w>\frac{\sqrt{13}-4}{3} \right)\\
  \Lambda<0,~~ \frac{3}{4 \Lambda}+\frac{\sqrt{13}}{4 |\Lambda|}<\alpha <-\frac{1}{6 \Lambda },~~
   \left(-\sqrt{\frac{4}{54 \alpha  \Lambda +9}}-\frac{1}{3}<w<-\frac{4+\sqrt{13}}{3}
   ~\mbox{or}~ -\frac{4+\sqrt{13}}{3} <w\leq -\sqrt{\frac{26 \alpha\Lambda +3}{18\alpha\Lambda }}-\frac{4}{3}~\mbox{or}~
   \right. \\ \left. \sqrt{\frac{26 \alpha  \Lambda +3}{18\alpha  \Lambda }}-\frac{4}{3}\leq w<\frac{\sqrt{13}-4}{3}~\mbox{or}~
   w>\frac{\sqrt{13}-4}{3}\right)\\
    \Lambda<0,~~ \alpha \geq -\frac{1}{6 \Lambda },~~ \left(w<-\frac{4+\sqrt{13}}{3}~\mbox{or} ~ -\frac{4+\sqrt{13}}{3}
   <w\leq -\sqrt{\frac{26 \alpha  \Lambda +3}{18\alpha  \Lambda }}-\frac{4}{3}~\mbox{or}~ \sqrt{\frac{26 \alpha  \Lambda +3}{18\alpha  \Lambda }}-\frac{4}{3}\leq w<\frac{\sqrt{13}-4}{3}~\mbox{or}\right. \\ \left.
    w>\frac{\sqrt{13}-4}{3}
  \right)
   \end{array}
\end{eqnarray}
and
\begin{eqnarray}\label{con2}
\begin{array}{ll}
\Lambda =0,~~ \alpha <0,~~ w>\frac{1}{3}\\
\Lambda=0,~~ \alpha >0,~ \left(-\frac{4+\sqrt{13}}{3}<w<-1 ~\mbox{or}~ -1<w<\frac{\sqrt{13}-4}{3} \right)\\
\Lambda >0,~~ -\frac{1}{6 \Lambda }<\alpha <0,~~ w> \sqrt{\frac{4}{54 \alpha  \Lambda
   +9}}-\frac{1}{3}\\
\Lambda>0,~~  \alpha >0,~~ \left(-\frac{4+\sqrt{13}}{3}<w<-1~\mbox{or}~ -\sqrt{\frac{4}{54 \alpha
   \Lambda +9}}-\frac{1}{3}<w<\frac{\sqrt{13}-4}{3}\right)\\
   \Lambda <0,~~ \frac{1}{2 \Lambda }<\alpha <0,~~ \sqrt{\frac{4}{54 \alpha  \Lambda
   +9}}-\frac{1}{3}<w\leq \sqrt{\frac{26 \alpha  \Lambda +3}{18\alpha  \Lambda }}-\frac{4}{3}\\
   \Lambda <0,~~ 0<\alpha \leq -\frac{1}{8 \Lambda },~~ \left(-\frac{4+\sqrt{13}}{3}
   <w<-\sqrt{\frac{4}{54 \alpha  \Lambda +9}}-\frac{1}{3}~\mbox{or}~
   -1<w<\frac{\sqrt{13}-4}{3}\right)
 \end{array}
\end{eqnarray} respectively corresponding to the CPs $A_+,~A_-$. As a result, CPs $A_\pm$ can address the physical
stable ESUs in the context of EMSG model with a flat spacetime, if the above conditions are satisfied.

Let's go to
CPs $C_\pm$ and $D_\pm$ arising from the spatially non-flat cases. Here, by imposing the conditions $\rho_{ES}\geq0$ and
$\frac{1}{a_{ES}^2}>0$ along with the stability condition $\lambda^2<0$,
it is not possible to list constraints such as (\ref{con1}) and (\ref{con2}).
Instead we illustrate the parameter space plots
in terms of $w-\alpha$ which address the allowed regions for existence of real and stable ESUs, see Figs.~\ref{cdp}
and ~\ref{cdn}. As it is clear, the only CP which admits  three values for the cosmological
constant is $C_-$. The CPs $C_+$ and $D_+$ admit values $\Lambda<0$ and $\Lambda>0$ while for the CP $D_-$ just the value $\Lambda<0$ is allowed.
As an additional check, in the Fig.~\ref{pd} optionally for three CPs $A_{-},~C_+$ and $D_+$, we present the phase-space behavior for various set parameters $(w,\alpha,\Lambda)$.

An important issue has to be clarified at this point. The EMSG scenario modifies  GR in the early phase of the universe. According to this statement,   the  CPs    seem no longer  applicable in
classical evolution, because, in the  EMSG phase, the universe undergoes a quantum regime. In this situation, the  classical evolution and expansion seem forbidden. To address this concern, one has to  note the following discussion.
In \cite{Roshan:2016mbt}, it is   shown that, in the EMSG picture, by controlling the values of
 free parameter $\alpha$, the universe no longer enter a quantum era during its evolution. In  other words, to guarantee that the  CPs  are free from quantum effects and allow the classical evolution, one has to impose  additional conditions as $\rho_{ES}<\rho_{p}$ and $a_{ES}<\ell_p$ where $\rho_{p}$ and $\ell_p$ are the Planck energy density and  the Planck length, respectively. More precisely, in our set up, the above mentioned constraints read as $\frac{1}{a_{ES}^2}<8\pi$ and $\rho_{ES}<64\pi$ since $c,~\hbar,~\kappa=1$ and $G=\frac{1}{8\pi}$. All numerical values taken into account for the EMSG parameter $\alpha$ throughout this paper satisfy both constraints, indicating that our analysis is safely classic and there is no need for any quantum considerations on the CPs. According to this result, the classical evolution is allowed.

\subsection{Non-Singular oscillations around stable ESU(s)} \label{osci}
\begin{figure}[!ht]
	\begin{center}
			\includegraphics[scale=0.8]{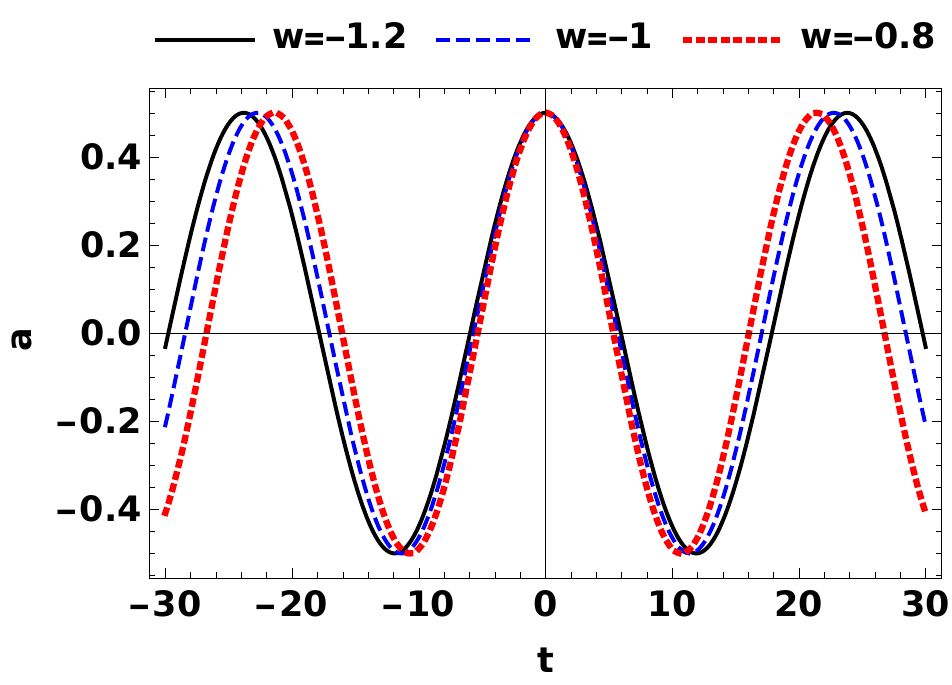}
			\includegraphics[scale=0.8]{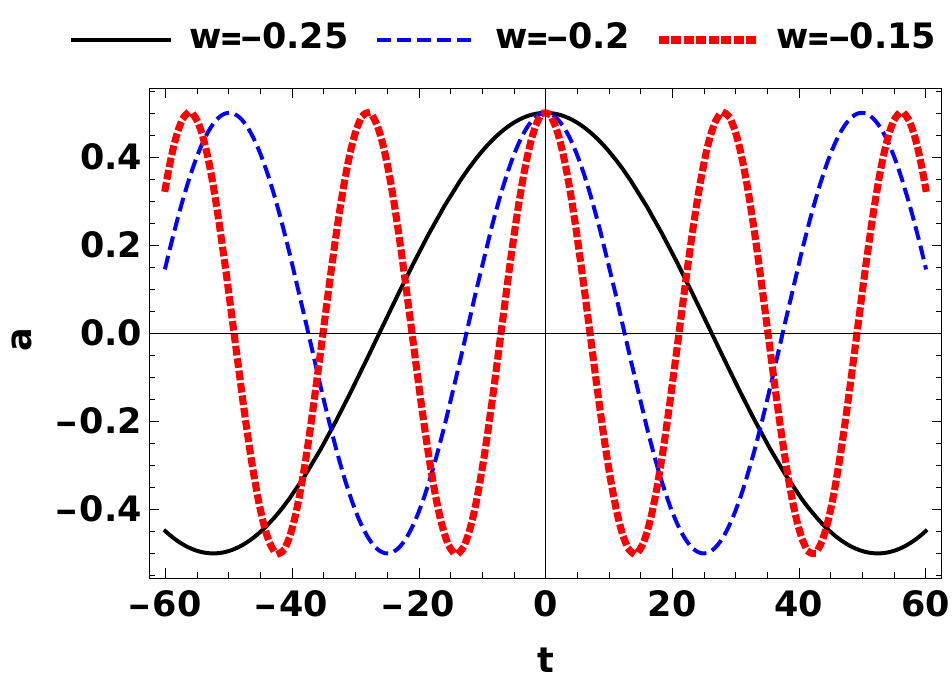}
            \caption{The evolutionary curves of scale factor corresponding to CPs $C_-$ (left panel) and $C_+$ (right panel)
            in terms of time with different values of $w$ and numerical values $\Lambda=0.5,~\alpha=-1$ and $\Lambda=0,~\alpha=-1$,
            respectively. }
		\label{OS1}
	\end{center}
\end{figure}

\begin{figure}[!ht]
	\begin{center}
			\includegraphics[scale=0.8]{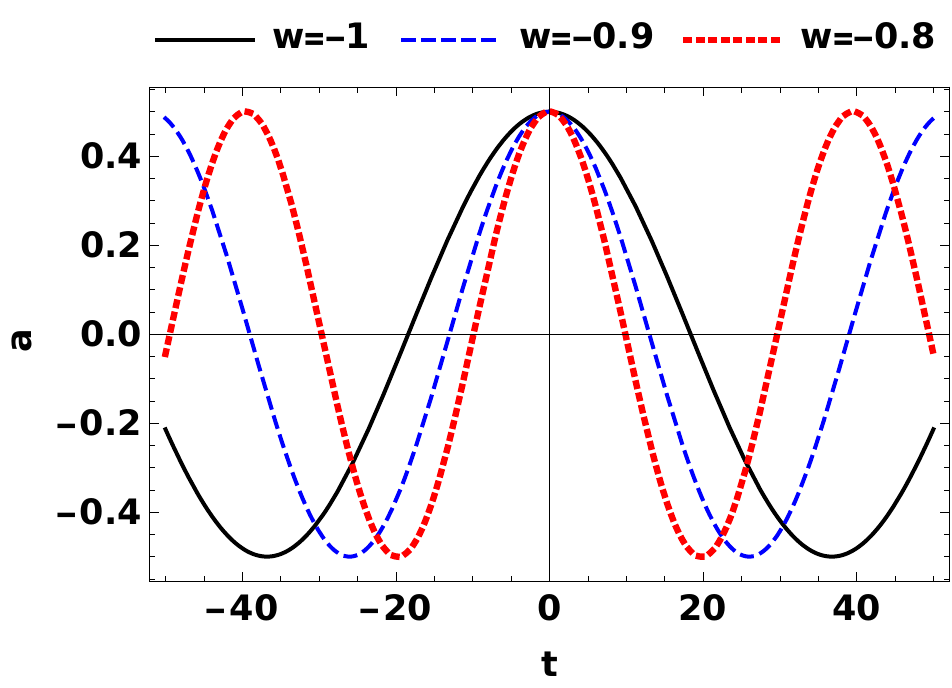}
			\includegraphics[scale=0.8]{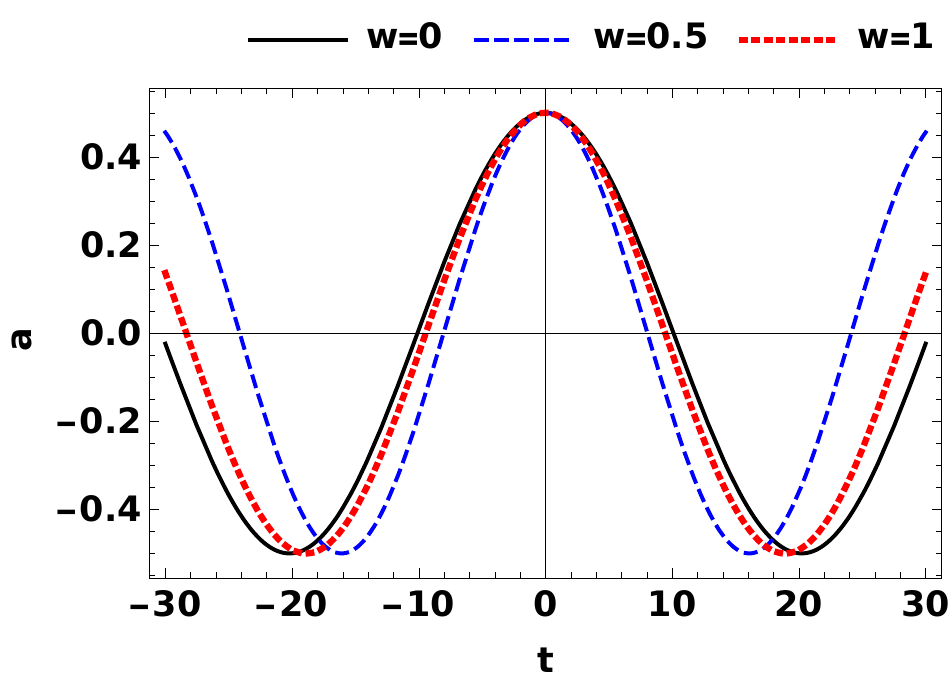}
            \caption{The evolutionary curves of scale factor corresponding to CPs $D_-$ (left panel) and $D_+$ (right panel)
            in terms of time with different values of $w$ and numerical values $\Lambda=-1,~\alpha=2$ and $\Lambda=-0.5,~\alpha=5$, respectively. }
		\label{OS2}
	\end{center}
\end{figure}

To ensure that the previous results do show stable points, we perform analysis of small perturbations around the ESU phases.
In what follows, we will show that if we  impose some small perturbations around the above mentioned stable CPs (in particular
spatially non-flat cases), the perturbations will undergo  an infinite series of non-singular oscillations around the inception
state. For this purpose, we linearly perturb Eqs. \eqref{frttfriedmannb} and \eqref{frttaccelb}
around the ESUs. We define the perturbations in the scale factor and matter density as follows
\begin{eqnarray}
\label{a}
&&a(t)\rightarrow a_{ES}(1+\delta a(t)),\nonumber\\
&&\rho(t)\rightarrow \rho_{ES}(1+\delta \rho(t))~,
\end{eqnarray} and also we have the linear expansions
\begin{eqnarray}
\label{b}
&&(1+\delta a(t))^{n}\simeq 1+n\delta a(t),\nonumber\\
&&(1+\delta \rho(t))^{n}\simeq 1+n\delta \rho(t).
\end{eqnarray}
After discarding the cross differential and non-linear terms, Eq. \eqref{frttfriedmannb} yields
\begin{eqnarray}\label{delta}
\frac{\delta \rho}{\delta a}=-\frac{6k}{a_{ES}^2(\rho_{ES}+2\alpha A_1)}~.
\end{eqnarray}
In the same way, using Eq.~\eqref{a} to perturb Eq. \eqref{frttaccelb} and
replacing the above relation, we acquire
\begin{eqnarray}\label{delta1}
\delta\ddot{a}-\bigg(\frac{(1+3w)k \rho_{ES}+4\alpha k A_2 \rho_{ES}^2}{a_{ES}^2(\rho_{ES}+2\alpha A_1)}\bigg)
\delta a=0~.
\end{eqnarray}
Now Eq. \eqref{delta1} allows us to consider the stability of the CPs extracted previously, via an alternative way i.e.
a small perturbation around CPs. As a cross-check by turning off the coupling parameter $\alpha$ in Eq.~\eqref{delta1},
we recover what is expected in the context of standard GR \cite{Barrow:2003ni}. In  Figs.~\ref{OS1} and~ \ref{OS2}, we represent the evolution of the scale factor
derived from Eq.~\eqref{delta1} corresponding to CPs  $C_{\pm}$ and $D_{\pm}$ with some optional parameters consistence with
Figs.~\ref{cdp} and~ \ref{cdn}. We observe the universe displays small oscillations around the relevant ESUs.

\begin{figure}[!ht]
	\begin{center}
		\includegraphics[scale=0.48]{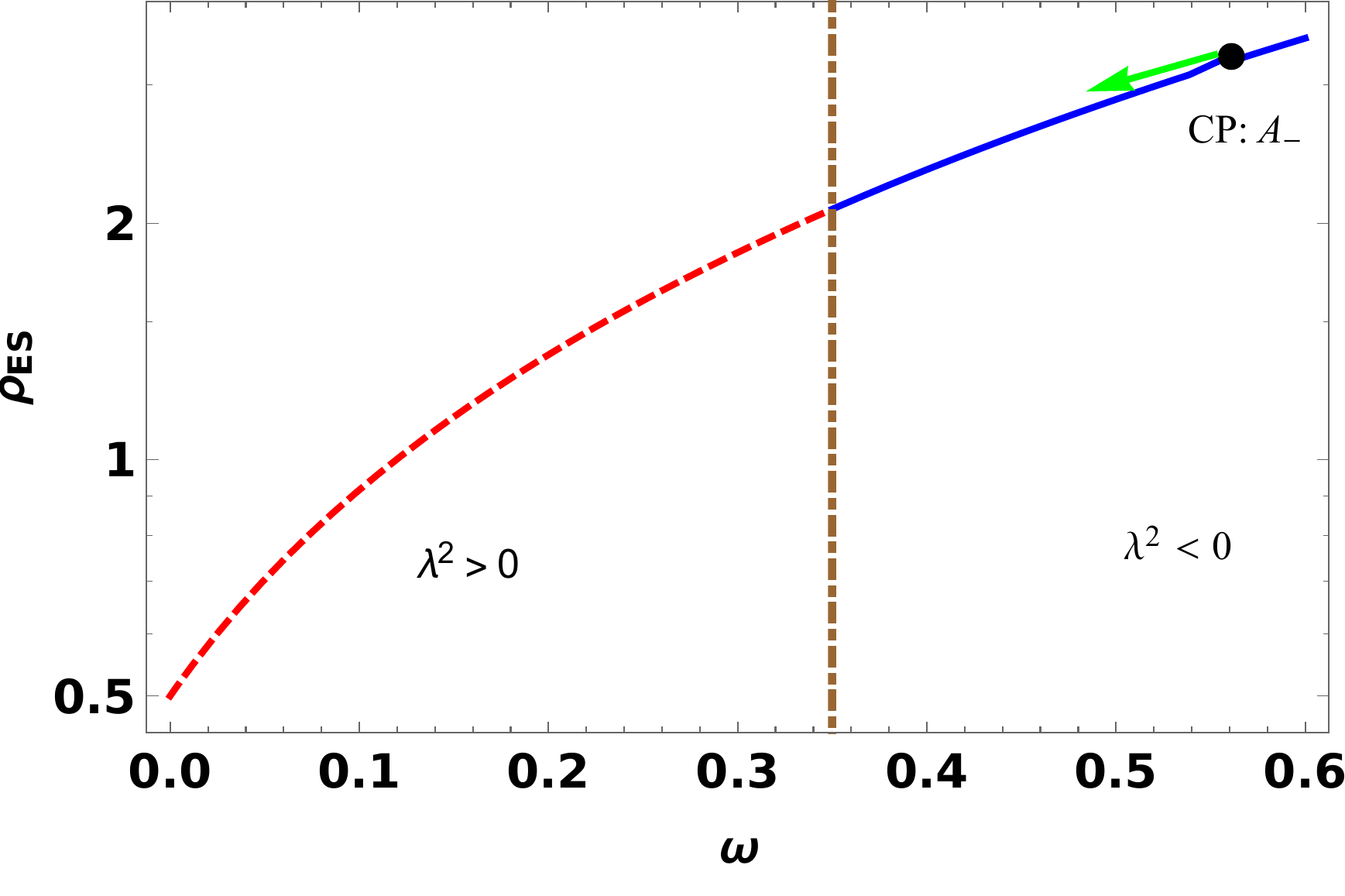}~~
		\includegraphics[scale=0.5]{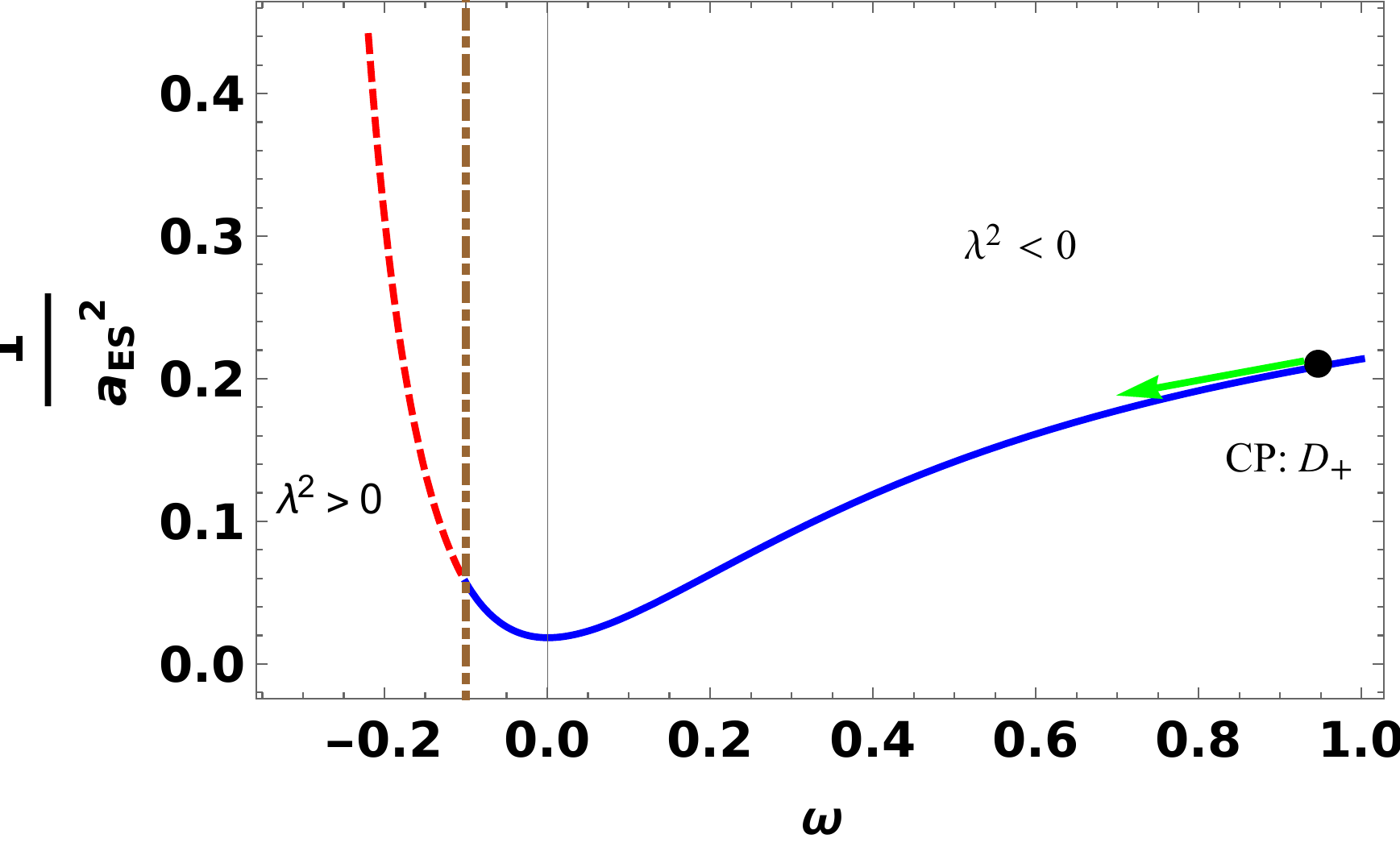}
		\caption{The general trend of transition of CPs $A_-$ corresponding to $\Lambda=0,~\alpha<0$ (left panel) and $D_+$ corresponding to $\Lambda<0,~\alpha>0$ (right panel) from stable region with $\lambda^2<0$ towards the unstable region with $\lambda^2>0$, as $\omega$ decreases. The vertical brown dashed-dotted line represents the critical value of $\omega$ where this transition occurs.}
		\label{TR}
	\end{center}
\end{figure}

\section{The realization of emergent cosmology scenario}\label{EC}
In this section, we intend to find a graceful exit mechanism for the stable
CPs in order to enter the standard thermal history of the universe by joining the CPs to
the inflationary period. More precisely, by  slowly reducing the universe
equation of state parameter $w$, we look for a phase transition from a stable state to
an unstable one for the aforementioned ESU solutions (CPs). In this way, we deal with a
non-singular early universe in a stable state  which finally will emerge
in a standard expanding universe.

We start with CPs derived for the spatially flat case i.e. $A_{\pm}$. To extract a stable to unstable
phase transition we need to know under what conditions the underlying CPs are both physical and  unstable.
Hence, for CPs $A_{\pm}$, we demand the conditions $\lambda^2 > 0$ as well as $\rho_{ES}\geq0$. This will yield the
following constraints on the set parameters $(w,\alpha,\Lambda)$
\begin{eqnarray}\label{con3}
\begin{array}{ll}
\Lambda<0,~~\alpha < \frac{3-\sqrt{13}}{4 \Lambda }~,~~\left(-\sqrt{\frac{26 \alpha  \Lambda +3}{18\alpha  \Lambda }}-\frac{4}{3}\leq w<-\frac{4+\sqrt{13}}{3} ~\mbox{or}~ -\frac{4+\sqrt{13}}{3}<w<-1 ~\mbox{or}~ \right. \\ \left.
\sqrt{\frac{4}{54\alpha  \Lambda +9}}-\frac{1}{3}<w<\frac{\sqrt{13}-4}{3}~\mbox{or}~
\frac{\sqrt{13}-4}{3}<w\leq \sqrt{\frac{26 \alpha  \Lambda +3}{18\alpha  \Lambda }}-\frac{4}{3}\right)\\
 \Lambda<0,~~\alpha =\frac{3}{4 \Lambda}-\frac{\sqrt{13}}{4 |\Lambda|},~~
   \left(-\sqrt{\frac{26 \alpha  \Lambda +3}{18\alpha  \Lambda }}-\frac{4}{3}\leq
   w<-\frac{4+\sqrt{13}}{3}~\mbox{or}~ -\frac{4+\sqrt{13}}{3}<w<-1 ~\mbox{or}~~\right. \\ \left.
    \frac{\sqrt{13}-4}{3}<w\leq \sqrt{\frac{26 \alpha  \Lambda +3}{18\alpha  \Lambda }}-\frac{4}{3}\right)\\
     \Lambda<0~,~~\frac{1}{2 \Lambda}<\alpha <0,~~ \left(-\sqrt{\frac{26 \alpha  \Lambda +3}{18\alpha  \Lambda }}-\frac{4}{3}\leq w<\frac{1}{3},~\mbox{or}~ -\frac{\sqrt{13}+4}{3}<w< -1 \right) \\
   \Lambda<0,~~0<\alpha \leq -\frac{3}{26 \Lambda },~~ \left(w<-\frac{\sqrt{13}+4}{3}~\mbox{or}~ -\frac{\sqrt{13}+4}{3}<w<-1\right) \\
     \Lambda<0,~~   -\frac{3}{26 \Lambda }<\alpha <-\frac{1}{8 \Lambda }~,~~ \left(w<-\frac{4+\sqrt{13}}{3}~\mbox{or}~
  -\frac{4+\sqrt{13}}{3}<w\leq -\sqrt{\frac{26 \alpha  \Lambda +3}{18\alpha  \Lambda }}-\frac{4}{3}~\mbox{or}~ \sqrt{\frac{26 \alpha  \Lambda +3}{18\alpha  \Lambda }}-\frac{4}{3}\leq w<-1\right)\\
    \Lambda<0,~~\alpha =-\frac{1}{8 \Lambda },~~ \left(w<-\frac{4+\sqrt{13}}{3}~\mbox{or}~ -\frac{4+\sqrt{13}}{3}
   <w<-\sqrt{\frac{26 \alpha  \Lambda +3}{18\alpha  \Lambda }}-\frac{4}{3}\right)\\
    \Lambda<0,~~  -\frac{1}{8 \Lambda }<\alpha <\frac{3}{4 \Lambda}+\frac{\sqrt{13}}{4 |\Lambda|},~~
   \left(w<-\frac{3}{4 \Lambda}+\frac{\sqrt{13}}{4 |\Lambda|}~\mbox{or}~ -\frac{4+\sqrt{13}}{3}<w<-
   \sqrt{\frac{4}{54\alpha  \Lambda +9}}-\frac{1}{3}\right)\\
   \Lambda<0,~~ \alpha =\frac{3}{4 \Lambda}+\frac{\sqrt{13}}{4 |\Lambda|},~~ w<-\frac{4+\sqrt{13}}{3}\\
   \Lambda<0,~~\frac{3}{4 \Lambda}+\frac{\sqrt{13}}{4 |\Lambda|}<\alpha <-\frac{1}{6 \Lambda },~~
   w<-\sqrt{\frac{4}{54 \alpha  \Lambda +9}}-\frac{1}{3}
    \end{array}
\end{eqnarray}
and
\begin{eqnarray}\label{con4}
\begin{array}{ll}
\Lambda =0,~~ \alpha <0,~~ \left(w<-\frac{4+\sqrt{13}}{3}~\mbox{or}~ \frac{\sqrt{13}-4}{3}
   <w<\frac{1}{3}\right)\\
\Lambda >0,~~ \alpha \leq -\frac{1}{6 \Lambda },~~ \left(w<-\frac{4+\sqrt{13}}{3}~\mbox{or}~ w>\frac{\sqrt{13}-4}{3}
   \right)\\
   \Lambda>0,~~  -\frac{1}{6 \Lambda }<\alpha <0,~~ \left(w<-\frac{4+\sqrt{13}}{3}~\mbox{or}~ \frac{\sqrt{13}-4}{3}
   <w<\sqrt{\frac{4}{54 \alpha  \Lambda +9}}-\frac{1}{3}\right)\\
   \Lambda>0,~~ \alpha >0,~~ -1<w<-\sqrt{\frac{4}{54 \alpha  \Lambda +9}}-\frac{1}{3}\\
    \Lambda <0,~~ \alpha <\frac{1}{2 \Lambda },~~ \left(-\sqrt{\frac{26 \alpha  \Lambda +3}{18\alpha  \Lambda }}-\frac{4}{3}\leq w<-\frac{4+\sqrt{13}}{3} ~\mbox{or}~ \frac{\sqrt{13}-4}{3}
   <w\leq \sqrt{\frac{26 \alpha  \Lambda +3}{18\alpha  \Lambda }}-\frac{4}{3}\right)\\
    \Lambda <0,~~ \alpha =\frac{1}{2 \Lambda },~~ \left(-\sqrt{\frac{26 \alpha  \Lambda +3}{18\alpha  \Lambda }}
    -\frac{4}{3}\leq w<-\frac{4+\sqrt{13}}{3} ~\mbox{or}~ \frac{\sqrt{13}-4}{3}
   <w<\sqrt{\frac{26 \alpha  \Lambda +3}{18\alpha  \Lambda }}-\frac{4}{3}\right)\\
   \Lambda<0,~~\frac{1}{2 \Lambda }<\alpha <0,~~ \left(-\sqrt{\frac{26 \alpha \Lambda +3}{18\alpha \Lambda }}-\frac{4}{3}\leq w<-\frac{4+\sqrt{13}}{3}~\mbox{or}~  \frac{\sqrt{13}-4}{3}<w<\sqrt{\frac{4}{54 \alpha \Lambda +9}}-\frac{1}{3}\right)\\
   \Lambda<0,~~0<\alpha \leq -\frac{3}{26 \Lambda },~~ - \sqrt{\frac{4}{54 \alpha\Lambda
   +9}}-\frac{1}{3}<w<-1\\
   \Lambda<0,~~-\frac{3}{26 \Lambda }<\alpha <-\frac{1}{8 \Lambda },~~ \left(-\sqrt{\frac{4}{54 \alpha
   \Lambda +9}}-\frac{1}{3}<w\leq -\sqrt{\frac{26 +3}{18\alpha  \Lambda }}-\frac{4}{3}~\mbox{or}~ \sqrt{\frac{26 \alpha
   \Lambda +3}{18\alpha  \Lambda }}-\frac{4}{3}\leq w<-1\right)
 \end{array}
\end{eqnarray} respectively.
\begin{figure}[!ht]
	\begin{center}
			\includegraphics[scale=0.5]{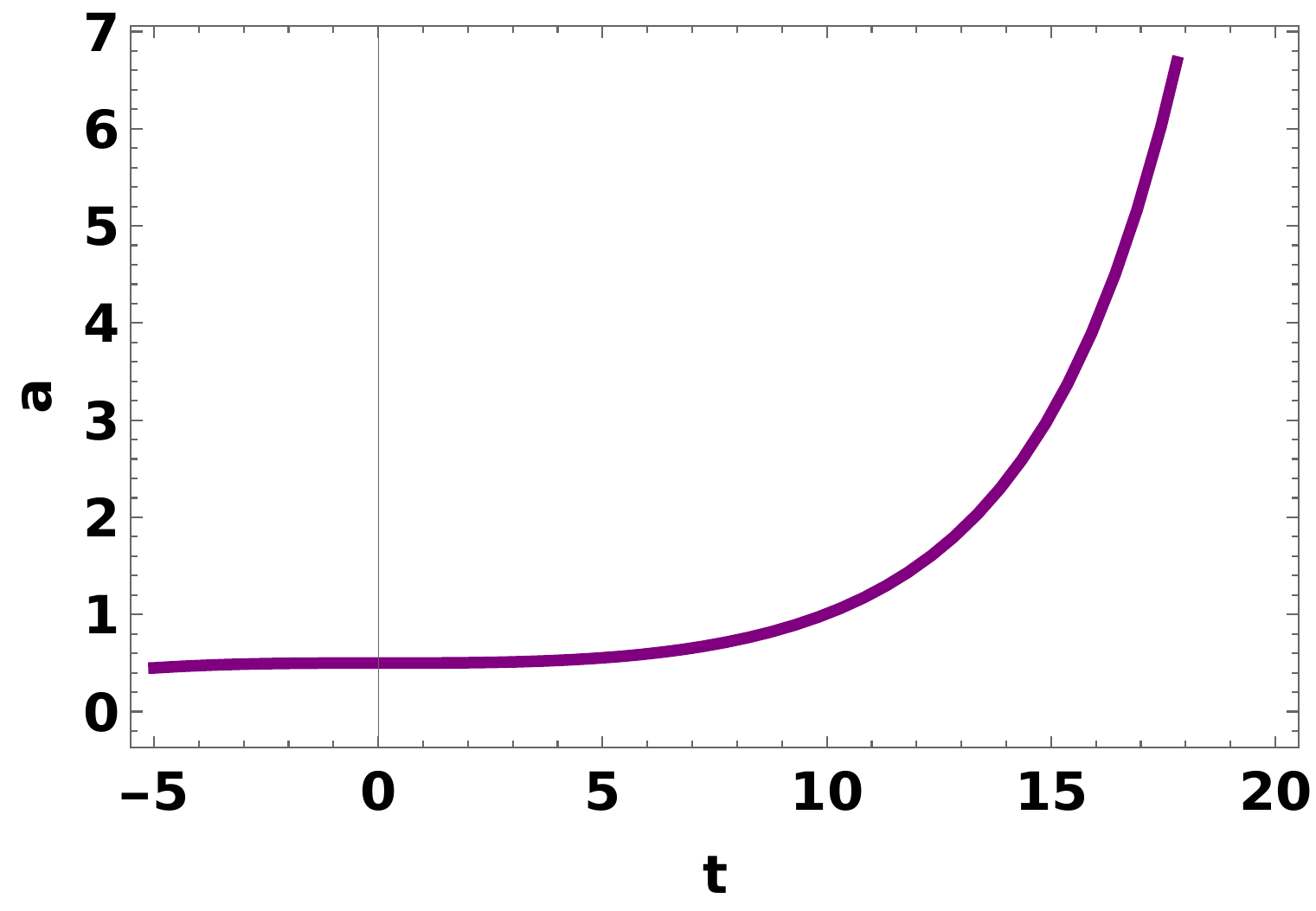}
			\includegraphics[scale=0.5]{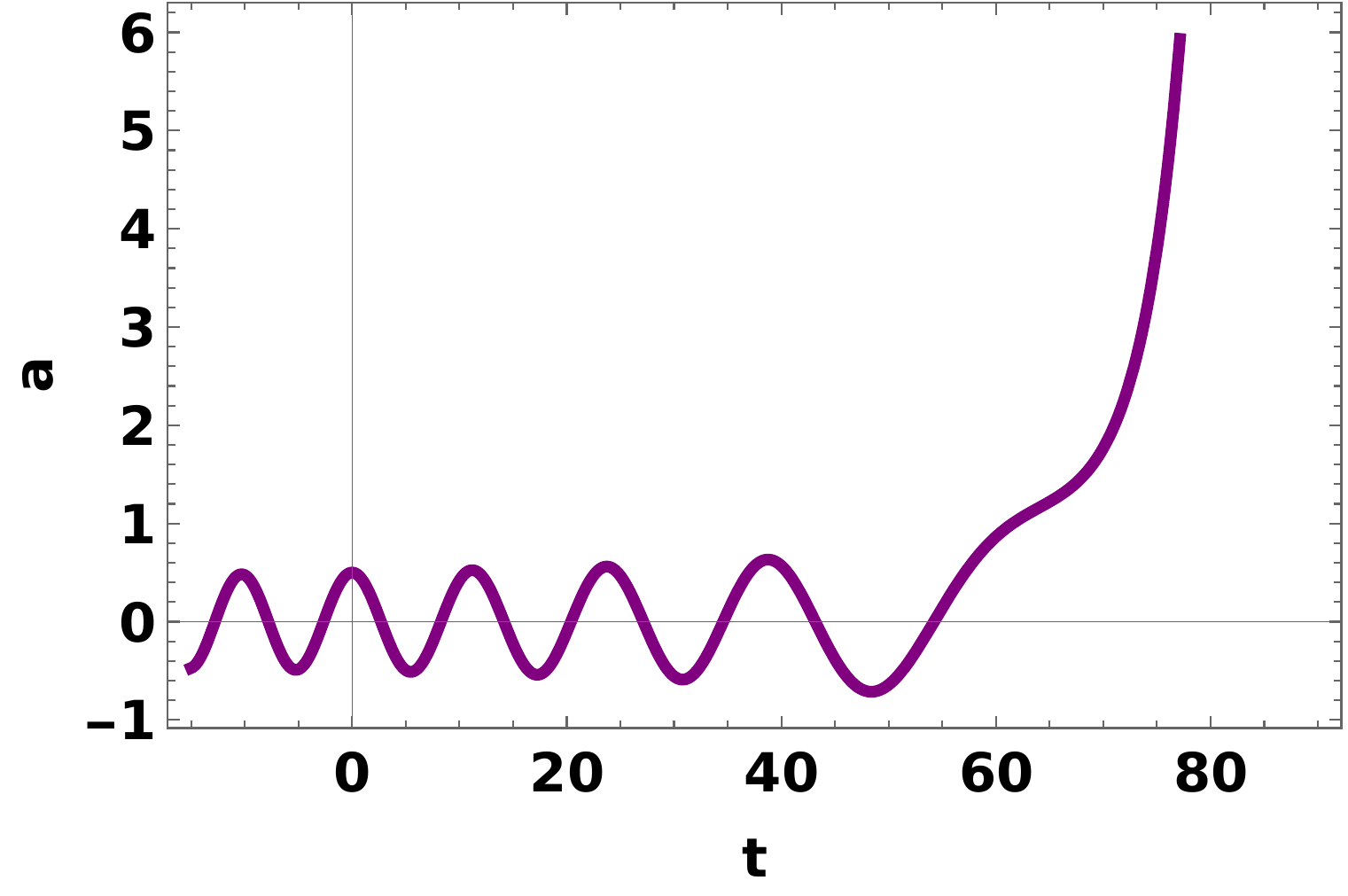}
            \caption{The graceful exit  of ESU from a stable state to an inflationary epoch by supposing a gradually
            decreasing  equation of state parameters  $w(t)=0.33-0.001t$ and $w=-0.3-0.01t$ for the EMSG model with flat
            and  spatially open geometries from left to right panels, respectively. We set numerical values ($\alpha=-1,
            ~\Lambda\geq0$) and ($\alpha=0.5,~\Lambda=-1$) with common initial conditions $a(0)=0.5$ and $\dot{a}(0)=0$ for
            left and right panels, respectively.}
		\label{FCO}
	\end{center}
     \end{figure}
     \begin{figure}[!ht]
	\begin{center}
			\includegraphics[scale=0.5]{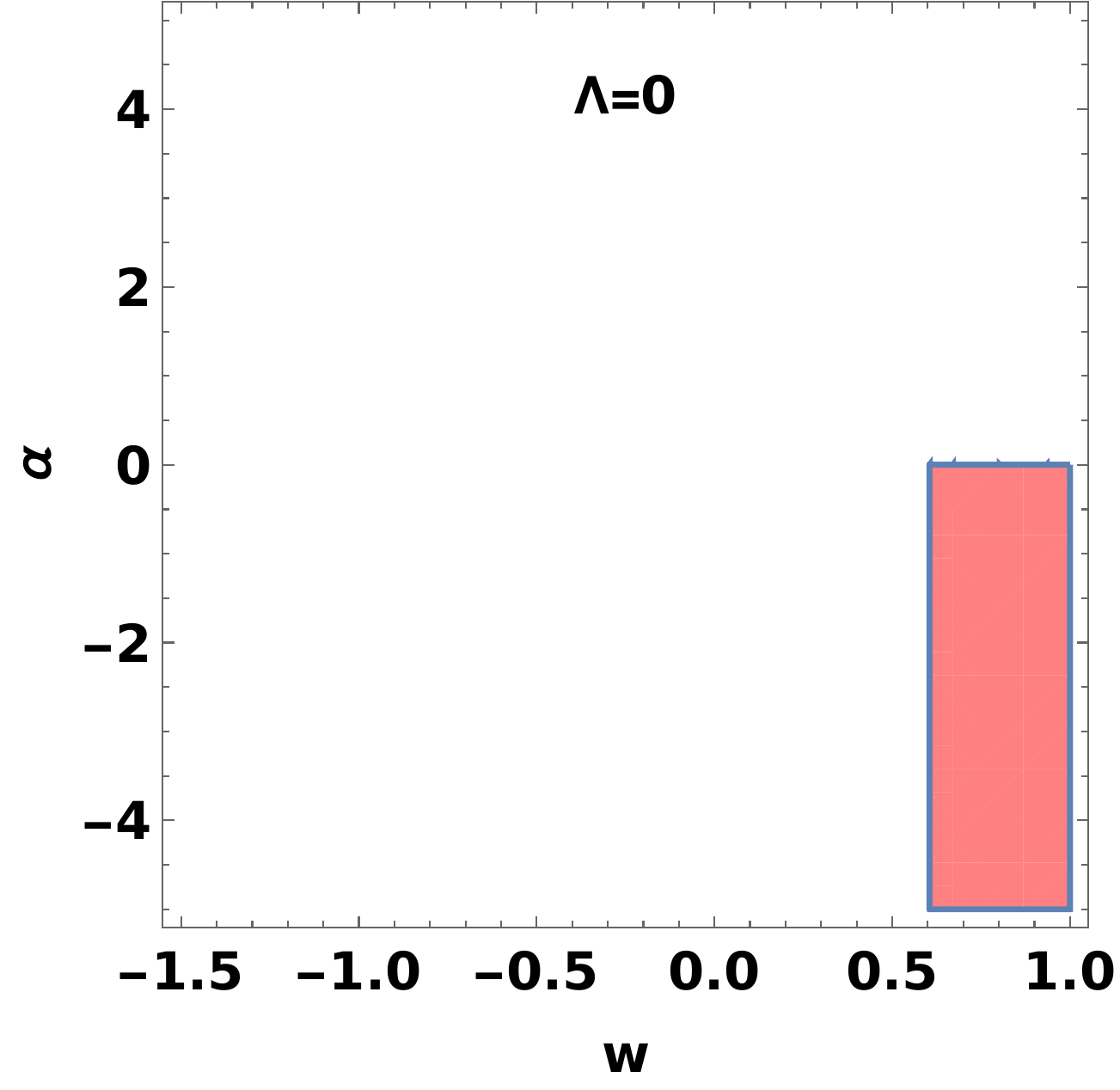}
			\includegraphics[scale=0.5]{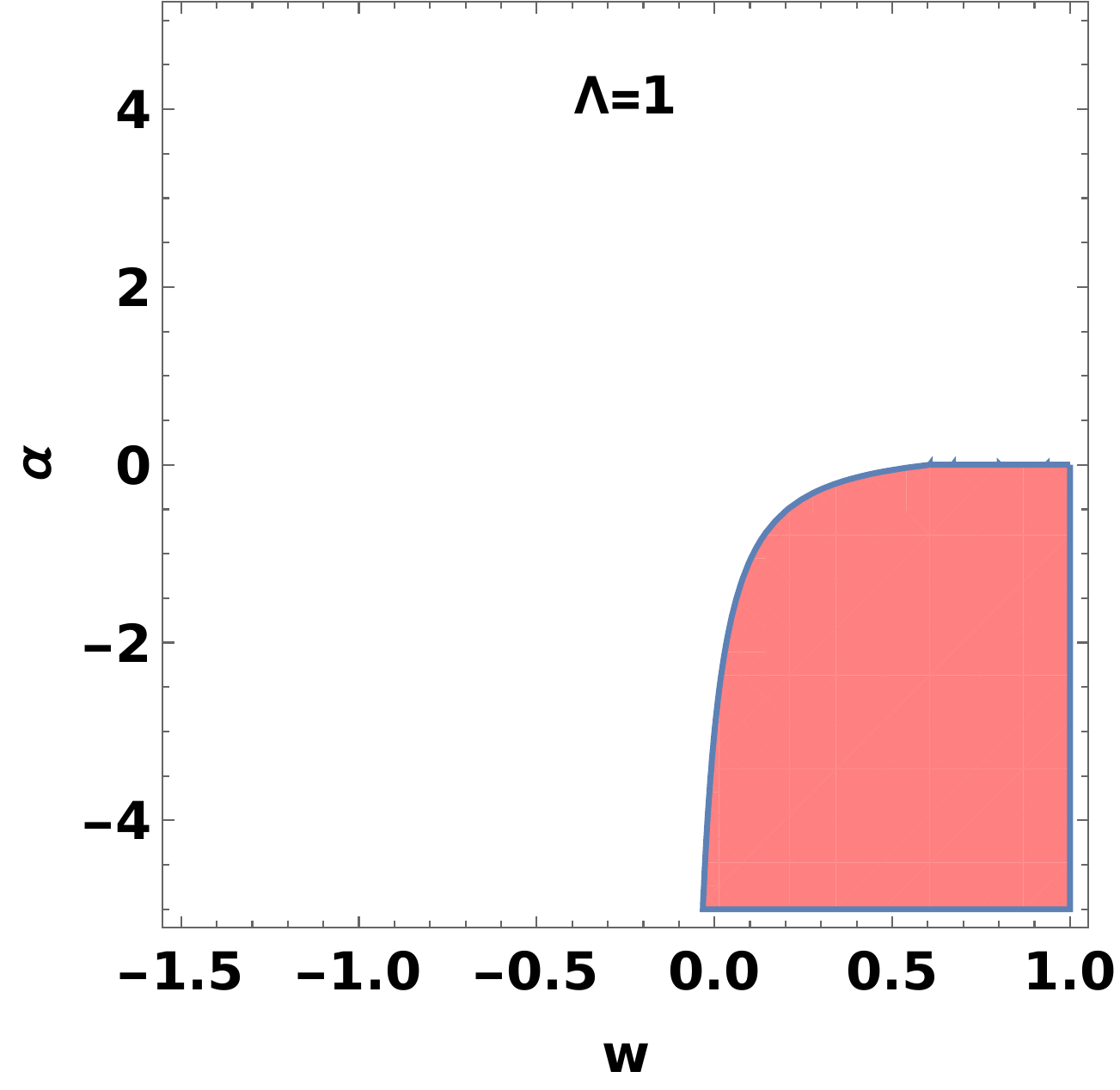}\\
            \includegraphics[scale=0.5]{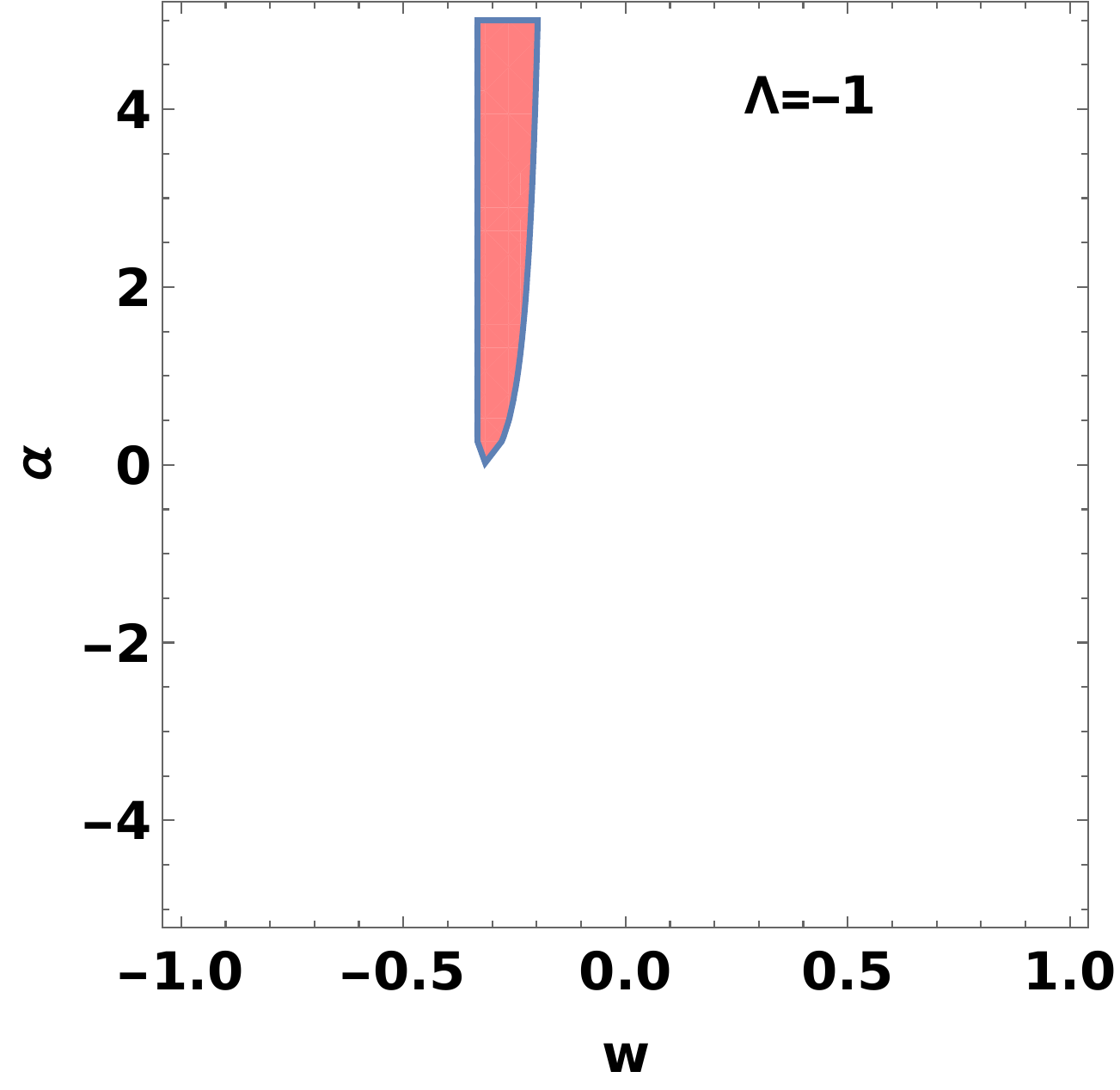}
			\includegraphics[scale=0.5]{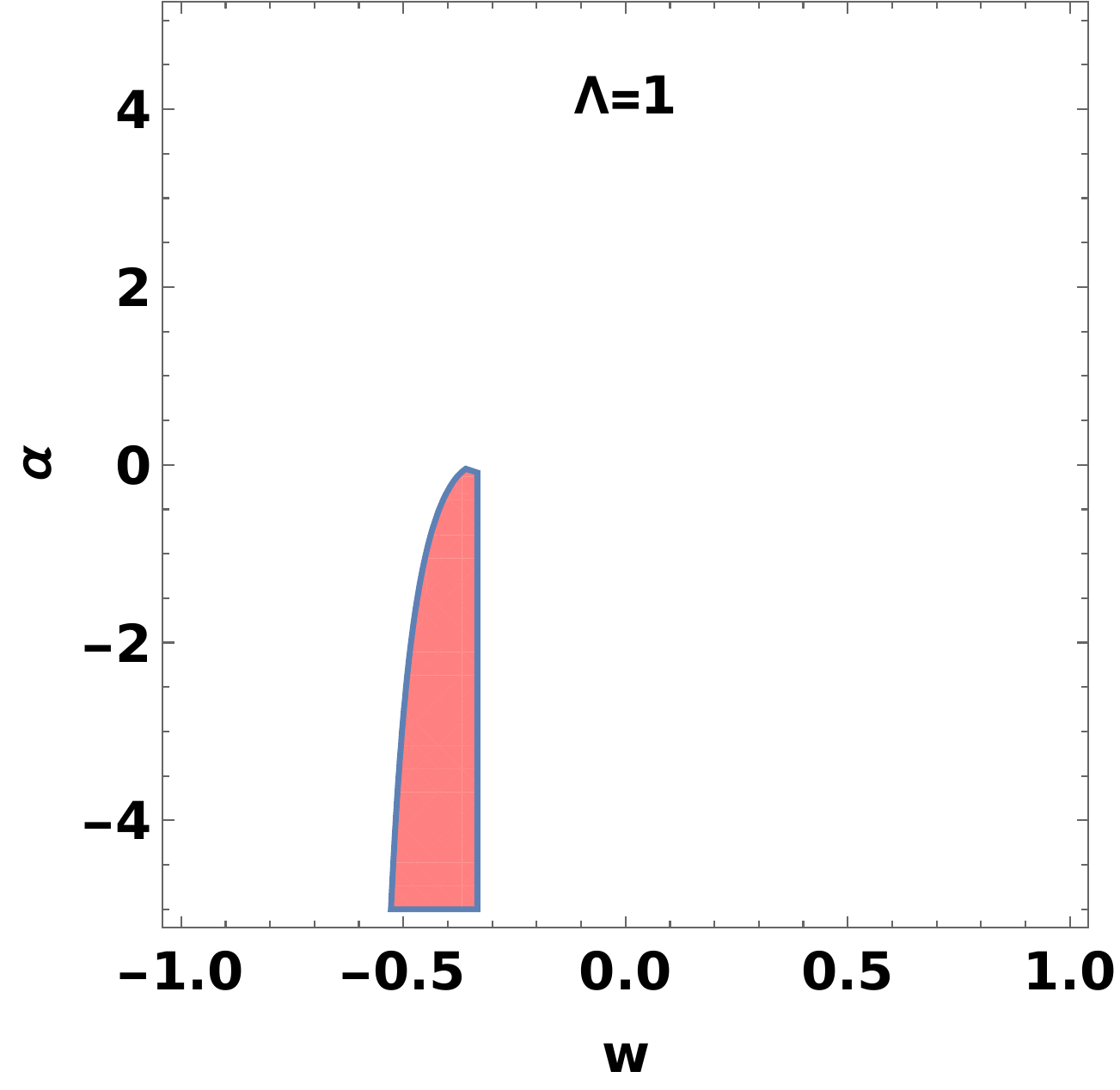}
            \caption{Regions of existence in the $(w,\alpha)$ parameter space for the unstable CPs $C_{-}$ (up row) and $D_{+}$
            (bottom row).}
		\label{UN}
	\end{center}
     \end{figure}
\begin{figure}[!ht]
	\begin{center}
			\includegraphics[scale=0.8]{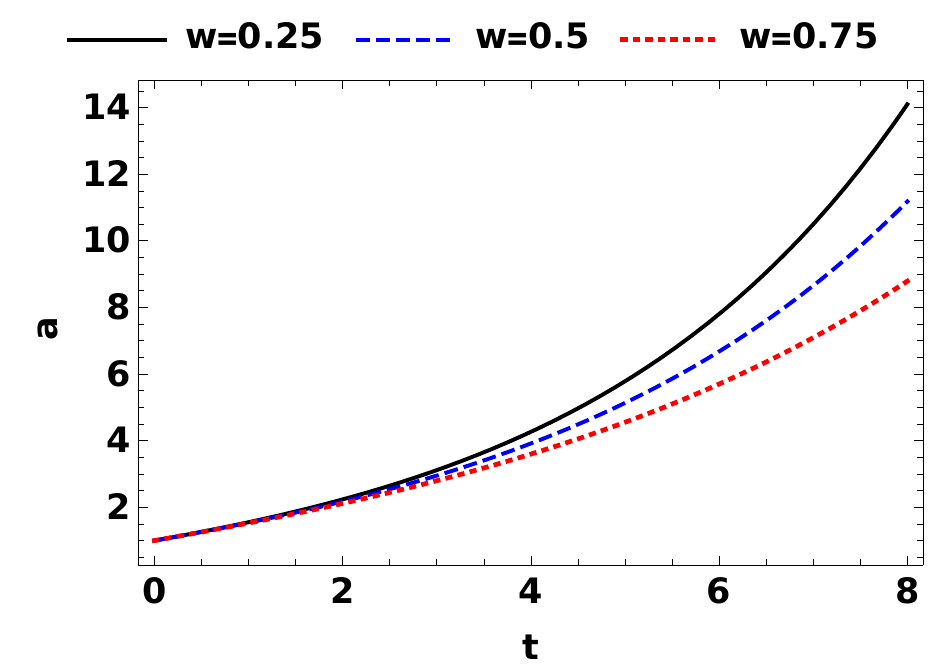}
			\includegraphics[scale=0.8]{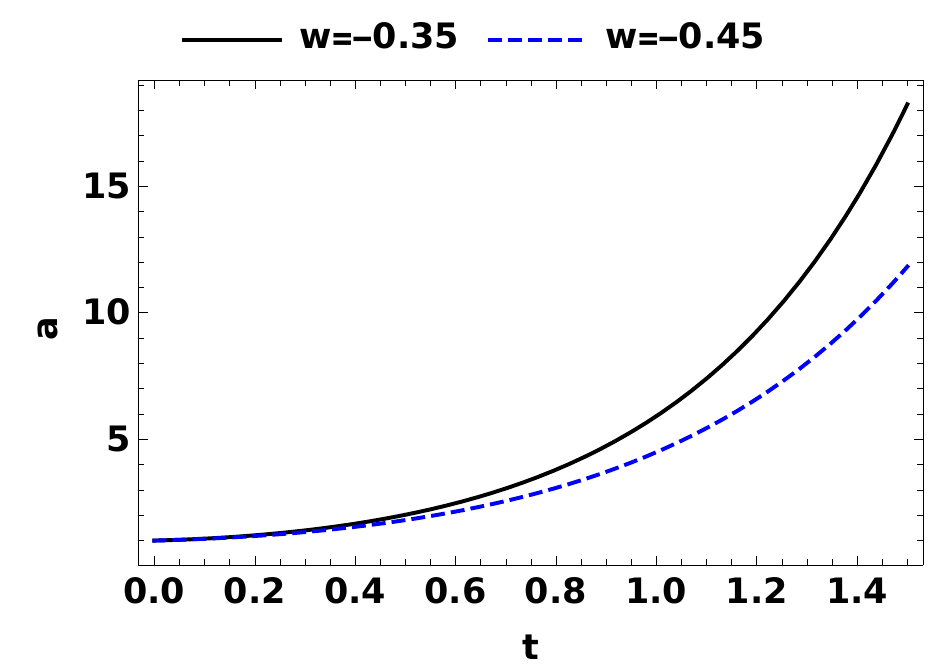}
            \caption{The increasing exponential behavior of the scale factor corresponding to the unstable CPs $C_{-}$ (left panel) and $D_{+}$ (right panel)
           with different values of $\omega$ and other numerical values related to Fig.~\ref{UN}. }
		\label{AI}
	\end{center}
     \end{figure}
Concerning the realization of the emergent universe scenario via the exit from the ESU,
the CPs $A_{\pm}$ show some promising behavior.

In general, we expect this transition to be achieved in two ways, as
the universe equation of state parameter slowly drops from a constant value in past ($t\rightarrow-\infty$).
Either via changing the stability status of each of CPs separately
from stable to unstable  or exchanging a stable CP with its unstable counterpart  so that
finally we will deal with an unstable CP \cite{Wu:2009ah, Zhang:2013ykz, Huang:2015zma, Khodadi:2015fav, Khodadi:2016gyw}.
Note that in both ways, it is expected that the phase transition occurs for a critical value of $\omega$.

Concerning the former, one can find from terms  in Eqs.~\eqref{con1} and ~\eqref{con3} that for the CP $A_{+}$ with negative
cosmological constant and negative values for the modified parameter $\alpha$, there is the
possibility of  the transition discussed above. By a closer inspection after comparing the first cases in Eqs.~\eqref{con1} and \eqref{con3}, we note that by setting values $\Lambda=-1,~\alpha=-1$, for spatially flat geometry, we can achieve
a stable ESU filled by a  matter with the equation of state parameter $w$ around $-1$. However, the CP $A_{+}$ is no longer stable after gradually dropping $w$ below
$-1$,  meaning that it undergoes a stable-unstable phase transition. Comparing the third cases in Eqs.~\eqref{con1} and \eqref{con3},  we can also have a stable ESU, but this time filled by a matter with the equation of state parameter
$w$ around $0$. After gradually decreasing $w$ below $0$, the CP $A_{+}$ exits from its stable status (for example, by setting values $\Lambda=-1,~\alpha=-1/4$). It turns out that this does not happen for positive values of $\alpha$.

Concerning CP $A_{-}$, we see from
comparison of the first cases in Eqs.~\eqref{con2} and \eqref{con4} that if $\Lambda=0,~\alpha<0$ then the phase transition does
occur for a ESU filled by  normal matter with the equation of state parameter $w$ around $1/3$. The third cases in  Eqs.~\eqref{con2}
and ~\eqref{con4} along with fifth item in Eq.~\eqref{con2} and the seventh in Eq.~\ref{con4}, signal the relevant phase transition for
$\Lambda>0$ and $\Lambda<0$ but with $\alpha<0$, respectively. Given the admitted non-negative values for the cosmological constant
by observation,  the CPs relevant to $\Lambda\geq0$ with $\alpha<0$ seem to be favored.
Besides, the constraint obtained from the neutron stars on the coupling parameter $\alpha$ implies that the EMSG model with $\alpha<0$, also has the potential to adapt with observation \cite{Akarsu:2018zxl}. Also, newly  in the light of observational data analysis of cosmic chronometers and Supernovae Type Ia data, shown that the flat EMSG model with $\alpha<0$ result in a favor and consistency cosmology in the absence of cosmological constant \cite{Ranjit:2020syg}. In \cite{Roshan:2016mbt}, also by setting the opposite sign convention to action \eqref{action} i.e. $R-\alpha T_{\mu\nu}T^{\mu\nu}$, it is clearly discussed that in de Sitter universe with a flat geometry to have a well-behaved cosmology in late-time which is stable, it is required that
$\alpha<0$.  As a result of this discussion, the cases $\Lambda\geq0,~\alpha<0$ are favored. In the left panel of Fig. ~\ref{TR}, we schematically show the general trend of the transition of CP $A_-$, corresponding to $\Lambda=0,~\alpha<0$, from  a stable region to an unstable one, as $\omega$ decreases and reaches a critical value.

Now we focus on the latter possibility for realization of phase transition. Comparing the third item in Eq.~\eqref{con1} with
the seventh in Eq.~\eqref{con4}, one finds that by setting values $\Lambda<0,~\frac{1}{2 \Lambda }<\alpha <0$ for spatially
flat geometry we can have a stable ESU addressed by CP $A_{+}$  filled by a matter with the equation of state parameter
$w$ around $-1$. However, by gradually decreasing $w$ from  $-1$ then the stable CP $A_{+}$  is exchanged with the unstable CP $A_{-}$. This
describes a stable-unstable phase transition. The sixth and fifth (also sixth) items in Eqs.~\eqref{con2} and \eqref{con3} respectively show
that there is also the possibility of transition from stable CP $A_{-}$ to unstable CP $A_{+}$ for positive values of $\alpha$
but with $\Lambda<0$. If the an anti-de Sitter universe finds any support by evidence, this case with $\alpha>0$ can be favored.
Left panel in Fig.~\ref{FCO} qualitatively displays the graceful exit from a stable state to inflation for the most popular possible
case between mentioned above cases (i.e $\Lambda\geq0$ and $\alpha<0$).

Considering the non-flat spatial geometries, let us start with closed universe ($k=+1$). By demanding instability conditions $\lambda^2>0$
,~$\rho_{ES}\geq0$ and $\frac{1}{a_{ES}^2}>0$ at the same time for the CPs $C_{\pm}$,  we deal with allowed regions in
$(w,\alpha)$  parameter space (Fig.~\ref{UN}). First of all, it is required  that the scale factor no longer  oscillates but grows exponentially to reach the inflationary phase. This can be done by setting the numerical values related to the allowed regions of Fig.~\ref{UN} into the perturbed Eq.~\eqref{delta1}, as displayed in Fig.~\ref{AI}. 

Interestingly, we find that the CP $C_+$ with non of the values of a cosmological constant
 admits these conditions. However, this is not the case for the CP $C_-$ with $\Lambda\geq0$, as
one can see in Fig.~\ref{UN} (up row). By comparing these plots with its stable counterpart in Fig.~\ref{cdn} (up row), one
recognizes that the stable-unstable phase transition is not possible as a result of the gradual decline of the equation
of state parameter.
However, by going to the spatially open  geometry case, we realize that just for the CP $D_+$
the underlying phase transition will happen. More explicitly, the bottom row plots in Figs.~
\ref{cdp} and \ref{UN}, address a stable ESU full of normal matter filled with $w\geq0$ for case of $\Lambda=-1$ and $\alpha>0$ which with the  passage of time and decline
of $w$, it is finally converted to an unstable state describing a phase transition.
The right panel of Fig. ~\ref{TR}, schematically shows the general trend of a transition from a stable region to an unstable one, as $w$ decreases.  The right panel of Fig.~\ref{FCO} also qualitatively shows the graceful exit from a
stable state to inflation for the CP $D_+$, as  the time passes.

%%%%%%%%%%%%%%%%%%%%%%%%%%%%%%%%%%%%%%%%%%%%%%%%%%%%%%%%%
\section{ Discussion and Conclusions }\label{CD}
Aiming to circumvent the initial singularity issue in the FLRW cosmology, we investigate one of the scenarios
under the spotlight in recent years named the Emergent Universe  Scenario. Based on this scenario,  our
universe does not stem from an origin as a big bang singularity, but it comes from an Einstein
static state in an infinite past and finally  joins the inflationary period. The
original framework for implementing this scenario was GR. Because GR does not admit
a stable static universe, this led to  the failure of this scenario in the first steps.
Subsequent studies have shown that taking into account some mechanisms, such as effects of modified
gravity, quantum gravity and extra dimensions, could improve original results in favor of
the realization of  Emergent Universe Scenario.

In this paper, we have implemented the Emergent Universe Scenario considering a modification of
Einstein gravity known as Energy-Momentum Squared Gravity (EMSG) which is distinguished from its
standard counterpart by the correcting term $T_{\mu \nu }T^{\mu \nu }$ in the action, that is considering a self-interaction of stress-energy tensor. By applying this
theory into cosmology, some modified terms, addressed by the  coupling parameter $\alpha$, appear in the
Friedmann equations describing the dynamics of universe. These new terms  can affect the
original Emergent Universe Scenario.

As a   first step, we  started our analysis by extracting the static
critical points known as Einstein Static Universe as a central concept in the study of the Emergent Universe
Scenario. By employing a first order dynamical analysis within the phase space  parameterized by $\alpha$, we  found some stable static critical points for any three possible spatial geometries ($k=0,\pm1$) in
the absence and presence of cosmological constant $\Lambda>0~(<0)$.

In the next step, by finding
a phase transition from a stable  to an unstable state for these extracted static critical points,
we  investigated the graceful exist of the Einstein Static Universe superseded by an inflationary epoch.
More precisely, the main idea is that for $t\rightarrow-\infty$, the value of $\omega$ is constant
and the universe, in essence, has been eternally stuck in the stable Einstein static state   until
it finally comes out naturally from this state and evolves into an  inflationary period, if $\omega$ drops
slowly as time goes forward. It is worth noticing that  when we say the Einstein Static Universe is stable or unstable, the statement  is true only within a specified range of $w$. Besides, in the downward trend of $w$, at  times when the cosmic $w$  approaches to critical values, some stable static solutions find the chance to exit from the allowed range of $w$ and, in this way, they enter into an unstable  range of $w$, meaning that the standard  universe evolution takes place. This can be seen schematically in Fig.~\ref{TR} for the static solutions $A_-$ and $D_+$.
As a result, this transition from stable to unstable Einstein Static Universe can be imagined natural in the sense that it occurs in two separated ranges of $w$, as the parameter decreases according to  the time evolution.

Interestingly, we  found that,
in the context of EMSG, unlike the original idea in  the standard general relativity, the realization of
an emergent universe  scenario does not impose  a positive spatial curvature ($k=+1$). In the other words,
our analysis have revealed that here the possibility of having an emergent universe for a closed universe, is ruled out.
However, for both spatially
flat $k=0$ and open $k=-1$ cases, under some constraints on the parameters ($\Lambda, \omega, \alpha$), there is
the possibility of a perfect realization for the Emergent Universe  Scenario. From three aspects, the most impressive output is related to the spatially flat case with $\Lambda=0$ and $\alpha<0$.

First, the spatially flat universe seems to be favored by  cosmological data, also if
the new Planck measurements could question this statement according to the discussion in \cite{DiValentino:2019qzk,Vagnozzi:2020rcz}. We leave its confirmation/rejection to the next data in the future.

Second, the Emergent Universe  Scenario works in the absence of a cosmological constant. This fact could be relevant in view of solving the cosmological constant problem \cite{Weinberg:1988cp}.

Third, it seems that the negative values of free model parameter $\alpha$ is supported by some observational constraints derived in \cite{Akarsu:2018zxl,Ranjit:2020syg}.

Our analysis, meanwhile, have shown that,
in the spatially flat case with $\alpha>0$, there exists the possibility of a successful realization for
the emergent universe if $\Lambda<0$, which despite its major role in AdS/CFT correspondence and a fundamental theory such as string, however, it is not supported by the cosmological observations. According to our analysis, by  the Emergent Universe Scenario in the context of EMSG, it is  possible to  bypass the initial singularity considering some favorable conditions as well as in absence of  quantum corrections. Results of our analysis become  remarkable considering those reported in  \cite{Barbar:2019rfn} where it is  shown that the initial bounce,  expected from EMSG \cite{Roshan:2016mbt},  is not viable as a regular bounce. This means that  finding a bounce solution  is not a sufficient condition to remove the singularity.  On the other hand,  we have shown that it is possible via a full realization of the Emergent Universe Scenario.

A further comment is worth at this point. The analysis of Emergent Universe Scenario can become richer, if one takes into account  the presence of  scalar fields in the EMSG  Lagrangian density.  This is due to the fact that EMSG not only changes the gravitational sector also modifies dynamics of  all involved matter fields. So, by considering  scalar fields  in the Lagrangian density,  due to the  higher derivative terms  $T_{\mu\nu}T^{\mu\nu}$ in \eqref{action},  non-trivial effects come out in  the Emergent Universe Scenario.

\section*{Acknowledgments}

SC acknowledges the support of {\it Istituto Nazionale di Fisica Nucleare} (INFN) ({\it iniziative specifiche}  MOONLIGHT2 and  QGSKY). This paper is partially based upon work from COST action CA15117 (CANTATA), and COST action CA18108 (QG-MM), supported by COST (European Cooperation in Science and Technology).

\end{document}